\documentclass[manuscript,screen]{acmart}
\usepackage{algorithm}
\usepackage{algorithmic}
\usepackage{bibentry}
\usepackage{amsfonts}
\usepackage{subfigure}
\usepackage{multirow}
\usepackage{amsmath}

\usepackage{graphicx}
\usepackage{epsfig}
\usepackage{color}
\usepackage{epstopdf}
\usepackage{rotating}
\usepackage{caption}
\usepackage{float}
\usepackage{multicol}
\usepackage{amssymb}
\usepackage{graphicx}
\usepackage{bbding}
\usepackage{algorithm}
\usepackage{algorithmic}
\usepackage{balance}
\usepackage{microtype}
\usepackage{graphicx}
\usepackage{subfigure}
\usepackage{booktabs}
\usepackage{multirow}
\usepackage{amsmath}
\usepackage{mathtools}
\usepackage{etoolbox}
\usepackage{cases}
\usepackage{enumitem}
\usepackage{xcolor}
\usepackage{subfigure}
\usepackage{cases}
\usepackage{threeparttable}

\newcommand{\ourname}{{SSD4Rec}}

\usepackage{tikz}

\AtBeginDocument{%
  }

\setcopyright{acmlicensed}
\copyrightyear{2024}
\acmYear{2024}
\acmDOI{XXXXXXX.XXXXXXX}

\acmConference[Conference acronym 'XX]{Make sure to enter the correct
  conference title from your rights confirmation emai}{June 03--05,
  2024}{Woodstock, NY}
\acmISBN{978-1-4503-XXXX-X/18/06}

\begin{document}

\title{SSD4Rec: A Structured State Space Duality Model for Efficient Sequential Recommendation}

\author{Haohao Qu}
\email{haohao.qu@connect.polyu.hk}
\orcid{0000-0001-7129-8586}
\author{Yifeng Zhang}
\email{yifeng.zhang@connect.polyu.hk}
\orcid{0009-0009-1702-0327}
\authornote{The authors contribute equally to this paper. The order of authorship is arranged alphabetically by their last names. }
\affiliation{%
  \institution{The Hong Kong Polytechnic Univeristy}
  \city{Hong Kong}
  \country{Hong Kong}
}


\author{Liangbo Ning}
\email{liangbo1123.ning@connect.polyu.hk}
\orcid{0000-0001-6903-8996}
\affiliation{%
  \institution{The Hong Kong Polytechnic Univeristy}
  \city{Hong Kong}
  \country{Hong Kong}
}

\author{Wenqi Fan}
\authornote{Corresponding author: Wenqi Fan, Department of Computing (COMP) \& Department of Management and Marketing (MM). }
\email{wenqifan03@gmail.com}
\orcid{ }
\affiliation{%
  \institution{The Hong Kong Polytechnic Univeristy}
  \city{Hong Kong}
  \country{Hong Kong}
}

\author{Qing Li}
\email{qing-prof.li@polyu.edu.hk}
\orcid{0000-0003-3370-471X}
\affiliation{%
  \institution{The Hong Kong Polytechnic Univeristy}
  \city{Hong Kong}
  \country{Hong Kong}
}

\renewcommand{\shortauthors}{Qu et al.}

\begin{abstract}
Sequential recommendation methods are crucial in modern recommender systems for their remarkable capability to understand a user's changing interests based on past interactions.
However, a significant challenge faced by current methods (e.g., RNN- or Transformer-based models) is to effectively and efficiently capture users' preferences by modeling long behavior sequences, which impedes their various applications like short video platforms where user interactions are numerous.
Recently, an emerging architecture named \textbf{Mamba}, built on state space models (SSM) with efficient hardware-aware designs, has showcased the tremendous potential for sequence modeling, presenting a compelling avenue for addressing the challenge effectively.
Inspired by this, we propose a novel generic and efficient framework (\textbf{\ourname{}}) for sequential recommendations, which explores the seamless adaptation of Mamba for recommendations.
Specifically, SSD4Rec marks the variable- and long-length item sequences with sequence registers and processes the item representations with a novel Masked Bidirectional Structured State Space Duality block.
This not only allows for hardware-aware matrix multiplication but also empowers outstanding capabilities in variable-length and long-range sequence modeling.
Extensive evaluations on four benchmark datasets demonstrate that the proposed model achieves state-of-the-art performance while maintaining near-linear scalability with user sequence length. 

\end{abstract}




\maketitle

\section{Introduction}
\label{se:intro}
Recommender systems (RecSys) stand out as a crucial branch in the realm of data mining, offering an essential and powerful solution to alleviate information overload issues and enhance user experiences across a wide array of applications~\cite{zhao2024recommender,qu2024tokenrec,fan2023trustworthy}. 
As a prevalent task in practical applications, sequential recommendation strives to capture users' dynamic preferences and predict the next items that users are likely to interact with based on their historical behaviors, which have been widely used in various domains, such as e-commerce~\cite{singer2022sequential,wang2020time}, streaming media~\cite{guo2019streaming,rappaz2021recommendation}, and news recommendation~\cite{wu2022news}. 
Differing from conventional RecSys~\cite{fan2022graph,fan2019graph}, which generates recommendations solely based on the user-item interaction in a static fashion, the user's preferences are often dynamic, changing with time and influenced by various factors in the context of sequential recommendation tasks. 
For example, after purchasing an iPad, a user might continue to buy related Apple accessories such as the Apple Pencil.

To capture the sequential dynamic interests from the users' historical interactions, a surge of studies have been proposed to leverage the powerful capabilities of advanced deep learning architectures, including Recurrent Neural Networks (RNNs)~\cite{chen2018sequential,ye2020time,duan2023long} and Transformer~\cite{kang2018self,sun2019bert4rec} for sequential recommendations. 
For example, 
\citet{donkers2017sequential} propose a novel extension of Recurrent Neural Networks tailored for sequential recommendations.
\citet{sun2019bert4rec} proposes BERT4Rec, which introduces and leverages deep bidirectional self-attention based on Transformer for modeling user behavior sequences.
Despite their prevalence and effectiveness in sequential recommendations, there are still some limitations for existing methods based on various deep learning architectures. 
For example, RNN-based recommender systems process users' historical sequences step by step, requiring the completion of hidden state calculations from the previous time step before making predictions for the next time step, leading to significant time costs. 
However, recommendation systems, such as Amazon, often demand real-time recommendations, as excessive delays can diminish user experience and impact company revenue~\cite{huang2015tencentrec,ma2020temporal}. 
Another prevalent approach, transformer-based RecSys, due to their utilization of self-attention mechanisms, exhibits a quadratic growth in computational complexity as the input sequence length increases. 
This implies that such models face challenges in handling users who interact with a large number of items.


Recently, a promising architecture, structured state space models (SSMs)~\cite{qu2024survey, lu2024structured,gu2021efficiently}, have emerged to efficiently capture complex dependencies in sequential data, becoming a formidable competitor to existing popular deep neural networks.
As one of the most successful SSM variants, \textbf{Mamba}~\cite{gu2023mamba,mamba2} achieves comparable modeling capabilities to Transformers while maintaining linear scalability with sequence length by introducing a Structured State Space Model and corresponding hardware-computational algorithms. 
Inspired by the success of Mamba-based models for capturing sequential and contextual information in various domains~\cite{qu2024survey}, such as text summarization~\cite{bronnec2024locost,ren2024samba}, speech analysis~\cite{jiang2024dual,li2024spmamba}, and question-answering systems~\cite{he2024densemamba,lieber2024jamba}, it is appealing to transfer this success from other domains to sequential recommendations to address the aforementioned limitations of existing sequential RecSys. 
More specifically, Mamba proposes attention-like matrix multiplication based on the property of Structured State Space Duality (SSD)~\cite{mamba2} to allow SSM to compute parallelly, which can decrease the time consumption of recurrent operations in RNNs.
Moreover, Mamba introduces a Structured State Space Model that circumvents the need for self-attention mechanisms. 
With the growth in input length, Mamba maintains linear time complexity, significantly boosting the model's capabilities in processing long sequence data.
Thus, given their advantages, Mamba provides great opportunities to advance sequential recommendations. 

However, directly applying Mamba to sequential recommendation remains highly challenging, primarily due to two key hurdles. 
Due to varying lengths of users' historical sequences towards items, 
most existing sequential modeling methods need \emph{padding} or \emph{truncation} operations for sequences with different lengths in recommender systems, resulting in excessive computational overhead (padding) or the potential loss of critical information that could enhance recommendations (truncation), as illustrated in Figure~\ref{fig:adaptation}.
Thus, the first challenge is how to model the dynamic users' preference with variable- and lone-length historical interaction sequences.
Meanwhile, naive Mamba-based models employ a \emph{left-to-right uni-directional} modeling paradigm, which limits the power of hidden representations for items in historical sequences~\cite{sun2019bert4rec}. 
Specifically, each item can only encode information from preceding items, making it difficult to capture precise user preferences and achieve optimal performance in sequential recommendations. 
This is due to the fact that users' historical interactions might not follow a rigidly ordered sequence in practice. 
Thus, the second challenge is how to capture context from both preceding items and subsequent items within the users' interaction history. 


To address such challenges, in this paper, a Structured State-Space Duality-empowered recommendation framework (\textbf{\ourname{}}) is proposed based on the advanced Mamba architecture to capture the dynamics underlying variable- and long-length user interaction sequences effectively and efficiently.  
A novel strategy is introduced to encode the user's sequential interactions into a variable-length item sequence without requiring padding and truncation, thereby accommodating an unlimited input window. 
Meanwhile, a Masked Bidirectional State Space Duality (Bi-SSD) block based on the Mamba architecture is designed to swallow the numerous item representations and predict corresponding user preferences for proper recommendations.

Our major contributions are summarized as follows:
\begin{itemize} 

\item We seamlessly integrate the Mamba into the recommendation framework and propose a novel Structured State Space Duality-empowered sequential recommendation (\textbf{\ourname{})}, which utilizes the advantages of Mamba to effectively and efficiently capture the dynamic users' preferences and accurately generate next-item recommendations. 

\item We propose a novel strategy to encode the users' historical behavior with varying lengths and a Masked Bidirectional Structured State-Space Duality (Bi-SSD) block to capture contextual information, particularly the long-range dependencies, for user behavior sequence modeling.


\item We conducted extensive experiments on four commonly used datasets to empirically showcase the efficacy of \ourname{}, including its superior recommendation performance and its scalability in predicting the preferences of users with variable- and long-length interaction histories. 
Specifically, SSD4Rec is 68.47\% and 116.67\% faster than the representative attention-based and SSM-based methods (i.e., SASRec~\cite{kang2018self} and Mamba4Rec~\cite{liu2024mamba4rec}), respectively, when training to extract features on user interactions at the length of 200 with hidden state dimension of 256.
\end{itemize}


The remainder of this paper is structured as follows. 
Section \ref{sec:methodlogy} introduces the
proposed approach, which is evaluated in Section \ref{sec:Experiments}. 
Then, Section \ref{sec:relatedwork} summarizes the recent development of sequential recommendation and sequence modeling. Finally, conclusions are drawn in Section \ref{sec:conclusion}.

\section{The Proposed Method}
\label{sec:methodlogy}
This section will begin by formulating the problem of sequential recommendation.
Then, we provide an overview of the proposed \ourname{}, followed by a detailed explanation of each model component.
Finally, we will discuss the model training and inference of the proposed \ourname{}.

\begin{figure*}[t]
\centering
{\includegraphics[width=\textwidth]{{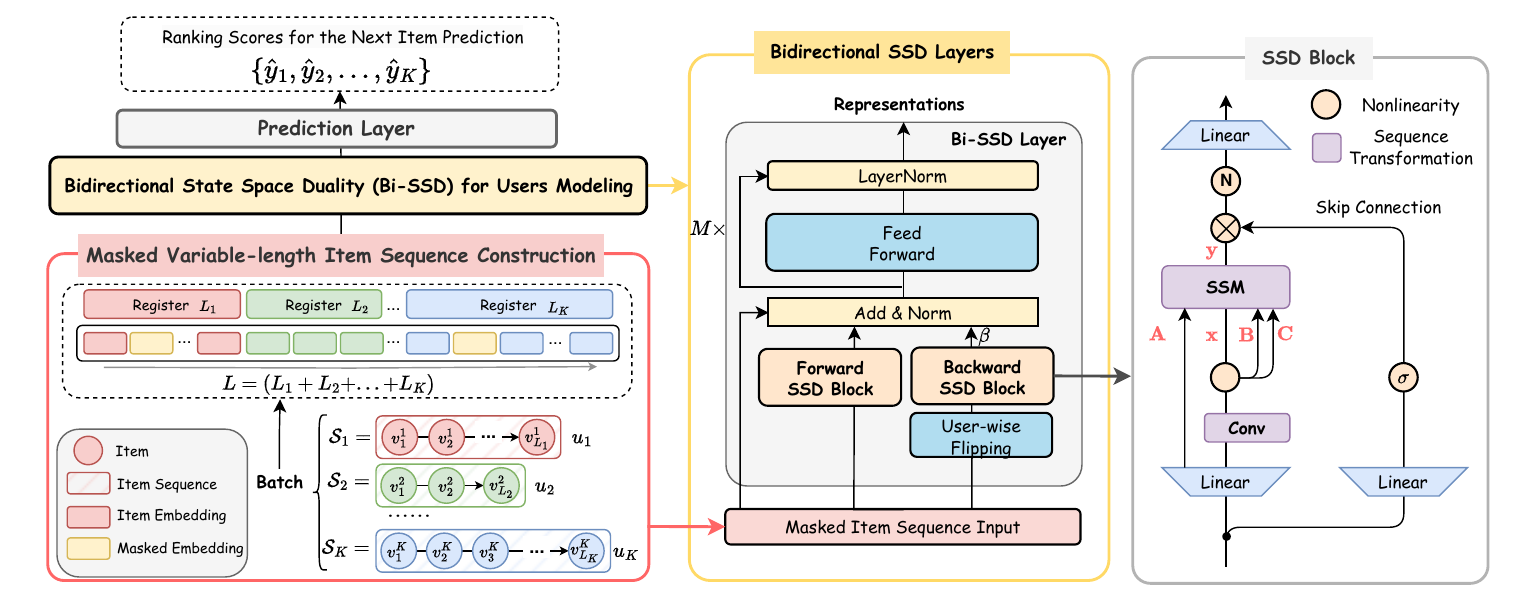}}}
\caption{The overall framework of the proposed \ourname{} for variable- and long-length sequential recommendation, 
which consists of the input construction with variable-length item sequences in a batch and the bidirectional block constructed with State Space Duality (SSD) layers for efficient and effective sequence modeling.} 
\label{fig:method}
\end{figure*}

\subsection{Notations and Definitions} \label{sec:task}
The primary goal of sequential recommendation is to predict the next interaction within a sequence. 
Let $\mathcal{U}=\{u_1,u_2,...,u_{|\mathcal{U}|}\}$ and $\mathcal{V}=\{v_1,v_2,...,v_{|\mathcal{V}}|\}$ be the sets of users and items, respectively. 
Each user $u_k \in \mathcal{U}$ is characterised by a chronologically ordered sequence $\mathcal{S}_k = [v^{u_k}_1, ..., v^{u_k}_t, ..., v^{u_k}_{L_k}]$, 
where $v^{u_k}_{t} \in \mathcal{V}$ represents the item that $u_k$ has interacted with at time step $t$, and $L_k > 1$ denotes the interaction sequence length for user $u_k$, which may vary from user to user.
Notably, for typical sequential recommendation methods, the input sequence length will be uniformed into a fixed hyper-parameter $L$ using the special padding token and truncation.
Specifically, if $L_k < L$, the sequence is modified to $\mathcal{S}_k^L = [pad, ..., pad, v^{u_k}_1, ..., v^{u_k}_{L_k}]$, with the number of padding instances equal to $(L - L_k)$. 
In contrast, if $L_k > L$, the sequence is truncated into $\mathcal{S}_k^L = [v^{u_k}_{(L_k-L)},...,v^{u_k}_{L_k}]$. 
To represent items efficiently, it is common practice to use an embedding table $\mathcal{Q} = \{\mathbf{q}_1, \mathbf{q}_2, ..., \mathbf{q}_{|\mathcal{V}|}\}$ for the item base $\mathcal{V}$, where $\mathbf{q}_j$ is a latent dense vector with dimensionality $D$ for item $v_j$. 

In general, a typical sequential recommendation method, denoted as $\textbf{SeqRec}$, analyzes the input sequence to capture user preferences and computes item ranking scores by performing the inner product between the predicted user preference and item embeddings.
For user $u_k$, such a process can be formulated as follows: 
\begin{align}
\label{eq:seqrec}
    \hat{\mathbf{p}}_k &= \textbf{SeqRec} (\mathcal{S}^L_k), \\
    r_k &= \hat{\mathbf{p}}_k \cdot \mathcal{Q},
\end{align}
where $\hat{\mathbf{p}}_k \in \mathbb{R}^D$ denotes the preference representation of user $u_k$, and $r_k$ is a score distribution of user $u_k$ over all items in the item base $\mathcal{Q}$.
This distribution is used to generate a ranked list of items representing the most likely interactions for user $u_k$ at step $L+1$.



\subsection{An Overview of the Proposed SSD4Rec}
Mamba models have recently demonstrated significant achievements across diverse domains, owing to their exceptional ability to capture intricate dynamics in long sequence modeling~\cite{gu2023mamba,mamba2}. 
By introducing advanced Structured State Space Models with hardware-aware expansions, Mamba facilitates attention-like modeling capabilities while maintaining a linear complexity in computing.
Motivated by these advancements, this paper proposes to explore Mamba's potential for effective and efficient sequential recommendations. 
Specifically, unlike the conventional sequential recommendation methods based on RNN or Transformer architecture, the proposed \ourname{} is designed to handle variable-length item sequences seamlessly without requiring padding or truncation, thereby ensuring that the model receives complete information and retains essential knowledge.
As shown in Figure~\ref{fig:method}, the proposed model consists of two key modules, namely \textbf{masked variable-length item sequence construction} and \textbf{Bidirectional Structured State Space Duality (Bi-SSD) Layer} for user modeling.
The first module aims to tackle the core challenge of managing variable-length sequences from diverse users within a certain batch.
Meanwhile, the proposed Bi-SSD layer enables attention-like content-aware interaction modeling while upholding linear complexity in computation.


\begin{figure}[t]
\centering
{\includegraphics[width=0.6\linewidth]{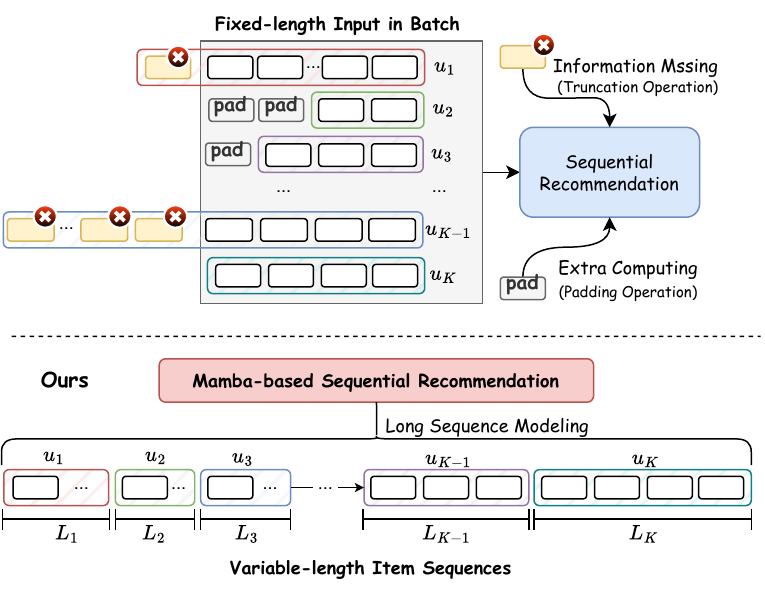}}
\caption{Compared to the typical sequential recommendation input with a fixed sequence length, the proposed input construction strategy achieves the seamless adaptation of Mamba to sequential recommendations, thus avoiding information loss and additional computation.
}
\label{fig:adaptation}
\end{figure}

\subsection{Masked Variable-length Item Sequence Construction}
The typical sequential recommendation methods (e.g., RNN-based and Transformer-based) perform padding or truncation operations to have fixed-length historical sequences of users' behaviors for the input of sequential architectures within a mini batch~\cite{kang2018self,chang2021sequential}. 
However, such fixed-length inputs suffer from intrinsic limitations, such as missing information and additional computational burdens for existing deep learning sequential recommendation methods, as depicted in Figure~\ref{fig:adaptation}. 
These limitations can lead to sub-optimal performance in capturing users' preferences for various modern web applications, such as e-commerce and social media platforms, where user interactions are significantly diverse. 
To address the problem, we introduce a novel strategy to encode user interactions of varying lengths within a mini-batch.
More specifically, within a batch, the proposed strategy combines item sequences of varying lengths into an extended integrated sequence and subsequently introduces \emph{segment registers} to facilitate user-aware interaction understanding.
Similar to the positional embeddings utilized in visual tasks for location-aware modeling, these registers delineate different segments of the integrated sequence originating from particular users.
By doing so, a standardized batch input can be built, encapsulating different item sequences from diverse users without the need for padding or truncation.


\subsubsection{\textbf{Masking Operation}}
It is well-proved that there are frequently unreliable user-item interactions~\cite{fan2022graph}, especially in long user behaviors, which can impede user modeling in existing data-driven models.
To enhance the robustness of \ourname{} against such noise, we propose to incorporate masking operations into interaction lists during the training phase. This approach creates a challenging task that helps the model develop a comprehensive understanding and generalize effectively. To be specific, we introduce an item-wise masking strategy $\gamma$ following Bernoulli distribution as follows:
\begin{align}
    \label{eq:bernoulli}
    \gamma \sim {\rm Bernoulli}(\rho),
\end{align}
where $\rho$ is the masking ratio. 
The Bernoulli distribution is the discrete probability distribution of a random variable, which takes the value 1 with probability $\rho$ and the value 0 with probability $(1-\rho)$.
Given the initial representations $\mathcal{E}_{u_k}=\{\mathbf{q}_1, \mathbf{q}_2, ..., \mathbf{q}_{L_k}\}^{u_k}$ of all the interacted items for user $u_k$, we replace the specific items determined by $\gamma$ with a masked embedding $\mathbf{m}$, as shown in Figure~\ref{fig:method}.
This masking operation randomly generates multifarious samples at each training epoch so as to enhance the generalization capability in the proposed model.

\subsubsection{\textbf{Variable-length Item Sequence}}

Formally, given $K$ users in the $b$-th batch $\mathcal{U}_b = \{u_1, ..., u_k, ..., u_K\}$ of varying interaction lengths, they can be characterized by their item sequences $\{\mathcal{S}_1, \mathcal{S}_k, ..., \mathcal{S}_K\}$, correspondingly.
In our case, the embeddings of these item sequences are concatenated into an extended integrated sequence, as expressed in
\begin{align}
    \label{eq:input}
   \mathcal{E}_b = [\underbrace{\mathcal{E}'_{u_1}, ..., \mathcal{E}'_{u_k}, ..., \mathcal{E}'_{u_K}}_{\text{$b$-th batch with size $K$}}],
\end{align}
where $\mathcal{E}'_{u_k} \in \mathbb{R}^{L_k \times D}$ denotes the masked sequence of item embeddings $\{\mathbf{q}_1, ..., \mathbf{q}_{L_k}\}^{u_k}$ for user $u_k$ with the sequence length of $L_k$, and $\mathcal{E}_b \in \mathbb{R}^{1 \times (\sum L) \times d}$ represents the integrated embedding sequence in the $b$-th batch. For simplicity, we omit the mark ($\prime$); that is, we still use $\mathcal{E}_{u_k}$ to denote the masked sequence embeddings for user $u_k$.
To delineate separate interactions, we incorporate segment registers based on user identifiers (IDs) with the integrated batch input.
The registers can be written by
\begin{align}
    \label{eq:register}
    \mathcal{R}_b = \{[u_1,...,u_1]^{L_1}, ... [u_k,...,u_k]^{L_k}, ..., [u_K, ..., u_K]^{L_K}\},
\end{align}
where $[u_k, ..., u_k]^{L_k}$ denotes the ID-based register of length $L_k$ for user $u_k$.
It is important to note that these registers do not possess a corresponding embedding; rather, they are employed to indicate the sequence computation to the SSD-based sequential recommendation.
As a result, the integrated sequence and its registers are combined and utilized as the model input $\mathcal{X}$ in the $b$-th batch:  
\begin{align}
    \mathcal{X} = (\mathcal{E}_b, \mathcal{R}_b).
\end{align}

In adopting this strategy, we transform the problem of batch processing sequences of varying lengths into a long sequence modeling task and leverage the capabilities of Mamba to address this issue effectively.
This strategy contributes to constructing a variable-length input for each batch without requiring padding or truncation, guaranteeing that the model receives comprehensive information from all user interactions.

\subsection{Bidirectional State Space Duality (Bi-SSD) for User Modeling}
With the substantial length of the constructed input $ \mathcal{X} = (\mathcal{E}_b, \mathcal{R}_b)$ and the diverse user information it contains, the primary challenge arises in the computation process.
In this context, RNN-based methods face challenges in training efficiency due to their lack of parallel computing capabilities, resulting in underutilization of the computational power offered by GPUs.
Moreover, as the sequence length expands, the complexity of Transformer-based methods escalates in quadratic time, leading to a notable increase in computing expenses. 
Thus, instead of processing all items in a recurrent format or calculating their attention scores, 
we propose to utilize a promising sequence modeling architecture, i.e., structured state-space models (SSMs), especially Mamba~\cite{gu2023mamba,mamba2}, to handle the lengthy input sequence for the sequential recommendation.
Specifically, inspired by the most advanced mamba framework based on state space duality (SSD) layers~\cite{mamba2}, our proposed \ourname{} is capable of executing block-based matrix multiplications on the extended input sequence $\mathcal{X}$, allowing segments associated with different users to be computed concurrently.
Furthermore, due to the complexities and variability of real-world behaviors, users’ historical interactions might not adhere to a strictly ordered sequence in practice~\cite{sun2019bert4rec}. 
However,  naive Mamba methods adhere to a left-to-right unidirectional modeling approach, constraining its modeling capabilities in such scenarios.
To this end, we propose to integrate reverse information from user interaction histories to construct a bidirectional block, enhancing Mamba's capacity for comprehensive recommendations.
In order to elucidate further, we will explain how \ourname{} processes the integrated input sequence leveraging the structured state-space duality property and detail the construction of the bidirectional block in the subsequent sub-sections.

\subsubsection{\textbf{Structured State Space Duality Block}}
To enable classical state space models the capabilities of content-aware modeling and efficient computation, Mamba~\cite{gu2023mamba,mamba2} introduces a selection mechanism to make state matrices become functions of model input and proposes hardware-aware algorithms based on the property of Structured State-Space Duality.
In general, a standardized SSM with structured state space duality processes an input $\mathbf{x} \in \mathbb{R}^{K \times L \times D}$ and its corresponding state matrices $\mathbf{A} \in \mathbb{R}^{K \times L \times D \times N}$, $\mathbf{B} \in \mathbb{R}^{K \times L \times N}$, and $\mathbf{C} \in \mathbb{R}^{K \times L \times N}$ to generate an output sequence $\mathbf{y} \in \mathbb{R}^{K \times L \times D}$,
where $K$, $L$, $D$, and $N$ represent the batch size, sequence length, input dimension, and hidden dimension.
This process can be expressed by
\begin{align}
    \label{eq:matrix}
    \mathbf{y} &= \textbf{SSM}(\mathbf{A}, \mathbf{B}, \mathbf{C}) (\mathbf{x}) = \mathbf{M} \mathbf{x}, \\
    \text{s.t.}~\mathbf{M} &= 
    \begin{bmatrix}
        (\mathbf{C}^0)^\top \mathbf{A}^{0:0} \mathbf{B}^0 & & & \\
        (\mathbf{C}^1)^\top \mathbf{A}^{1:0} \mathbf{B}^0 & (\mathbf{C}^1)^\top \mathbf{A}^{1:1} \mathbf{B}^1 & & \\
        ... & ... & ... &  \\
        (\mathbf{C}^j)^\top \mathbf{A}^{j:0} \mathbf{B}^0 & (\mathbf{C}^j)^\top \mathbf{A}^{j:1} \mathbf{B}^1 & ... & (\mathbf{C}^j)^\top \mathbf{A}^{j:i} \mathbf{B}^i \\
    \end{bmatrix},
\end{align}
where $\mathbf{M}^{ji} = (\mathbf{C}^j)^\top \mathbf{A}^{j:i} \mathbf{B}^i$ represents the mapping of the $i$-th token in the input sequence $\mathbf{x}$ to the $j$-th token in the output $\mathbf{y}$; ($\mathbf{A}^i$, $\mathbf{B}^i$, $\mathbf{C}^i$) denote the state matrices for the $i$-th token, derived through linear projection from the corresponding token embedding; and $\mathbf{A}^{j:i}$ is the product of the state matrix $\mathbf{A}$ from $j$ to $i$.
Given such a formulation, the structured state-space model allows for block-based matrix multiplication, thus achieving attention-like content-aware modeling and subquadratic-time computation. 

Building upon this insight, we further take advantage of the state space duality property for our variable- and long-length user interaction sequence. 
Unlike the standard formulation mentioned above, we propose to perform the block-based matrix multiplications concurrently for each user sequence in the integrated input $\mathcal{X}$. 
Thanks to the use of segment registers, the model can retrieve corresponding structured matrix $\mathbf{M}$ and item sequence $\mathcal{E} \in \mathbb{R}^{1 \times (\sum L) \times D}$ for computation. 
Formally, given an input $\mathcal{X} = (\mathcal{E}_b, \mathcal{R}_b)$ involving $K$ users in the $b$-th batch, the parallel SSD-based computation process can be formulated as 
\begin{align}
    \label{eq:ssd_all}
    \mathcal{F}(\mathcal{X}) = \textbf{SSM}(\mathbf{A}, \mathbf{B}, \mathbf{C}) (\mathcal{X}) &= (\mathbf{M}_1 \mathcal{E}_{u_1}, ..., \mathbf{M}_k \mathcal{E}_{u_k}, ..., \mathbf{M}_K \mathcal{E}_{u_K}), \\
    \text{s.t.}~~ \mathbf{M}_k^{ji} &= (\mathbf{C}_k^{j})^\top \mathbf{A}_k^{j} ... \mathbf{A}_k^{i+1} \mathbf{B}_k^{i},
\end{align}
where $\mathcal{F}(\mathcal{X})$ is the intermediate output of an SSD block. $\mathbf{M}_k$ and $\mathcal{E}_{u_k}$ represent the structured matrix and item sequence for user $u_k$.
By adopting this approach, \ourname{} implements batch processing of user interaction sequences with varying lengths.

\subsubsection{\textbf{Bi-SSD Layer}}
By leveraging the state space duality (SSD) block as the foundational component, we construct a network architecture tailored to address the sequential recommendation task, as illustrated in Figure~\ref{fig:method}. First, in addition to the standard sequential information concerning user behaviors, we propose integrating reverse insights from the interaction history by constructing a bidirectional architecture on top of the unidirectional SSD block. This bidirectional design encourages the model to learn to reason in a reverse manner, potentially capturing more valuable patterns within the interaction histories.
Second, the utilization of SSD blocks enables our model to facilitate hardware-aware computation, i.e., block-based matrix multiplication~\cite{mamba2}, thereby achieving attention-like content-aware modeling while preserving efficient learning.
Moreover, by stacking multiple bidirectional SSD layers, the proposed model can achieve a deeper network structure with an increased number of learnable parameters.
Layer-wise normalization and embedding dropout are further utilized between the stacked cells to improve the model's robustness and stabilize the learning process.
Finally, the proposed Bi-SSD layers predict the representations of user preferences in a specific batch, which are used to calculate the ranking score distributions for Top-K sequential recommendations.
Given a batch of masked item sequence inputs $\mathcal{X} = (\mathcal{E}_b, \mathcal{R}_b)$, the computation  procedures of our Bi-SSD layer can be expressed as follows: 

\begin{itemize}
    \item \textbf{Forward SSD Block.} As detailed in Equation (\ref{eq:ssd_all}), we utilize SSD block to process the masked item sequence input $\mathcal{X} = (\mathcal{E}_b, \mathcal{R}_b)$ as follows:
\begin{align}
    \label{eq:forward}
    \mathcal{F}_{\rm forward}(\mathcal{X}) = \textbf{SSM}(\mathbf{A}, \mathbf{B}, \mathbf{C}) (\mathcal{X}) &= (\mathbf{M}_1 \mathcal{E}_{u_1}, ..., \mathbf{M}_k \mathcal{E}_{u_k}, ..., \mathbf{M}_K \mathcal{E}_{u_K}).
\end{align}

\item \textbf{Backward SSD Block.} Before passing the input through the Backward SSD block, we first introduce the user-wise flipping function, which flips the entire masked input sequence from $\mathcal{X} = (\mathcal{E}_b, \mathcal{R}_b)$ to $\mathcal{X}^f = (\mathcal{E}^f_b, \mathcal{R}_b)$:
\begin{align}
    \label{eq:flip}
    \mathcal{E}_b = [\mathcal{E}_{u_1}, ..., \mathcal{E}_{u_k}, ..., \mathcal{E}_{u_K}] \rightarrow \mathcal{E}^f_b = [\underbrace{\mathcal{E}^f_{u_1}, ..., \mathcal{E}^f_{u_k}, ..., \mathcal{E}^f_{u_K}}_{\text{$b$-th batch with size $K$}}],
\end{align}
where $\mathcal{E}^f_{u_k}$ denotes the flipped user-item sequence corresponding for user $u_k$. The flipped input $\mathcal{X}^f$ will be processed by the Backward SSD Block as follows:
\begin{align}
    \label{eq:backward}
    \mathcal{F}_{\rm backward}(\mathcal{X}) = \textbf{SSM}(\mathbf{A}, \mathbf{B}, \mathbf{C}) (\mathcal{X}^f) &= (\mathbf{M}_1 \mathcal{E}^f_{u_1}, ..., \mathbf{M}_k \mathcal{E}^f_{u_k}, ..., \mathbf{M}_K \mathcal{E}^f_{u_K}).
\end{align}
Notably, to conserve computational resources and GPU memory usage, we can utilize a shared SSD block for the forward and backward modeling processes, which means the state matrices for the forward SSD block and the backward SSD block are identical.

\item \textbf{Add \& Norm. \& Residual Connection} Instead of merely summing the output from these three paths, we introduce a weighted indicator $\beta$ to regulate the integration of messages between the forward and backward sequence modeling.
As we may stack the Bi-SSD layers, it becomes important to utilize residual connections, which allow the embedding of the last visited item to be propagated to the final layer.
Finally, to improve robustness and prevent overfitting, a layer normalization is applied, followed by dropout:
\begin{align}
    \label{eq:bi}
    \mathcal{F}'(\mathcal{X}) &= \mathcal{F}_{\rm forward}(\mathcal{X}) + \beta * \mathcal{F}_{\rm backward}(\mathcal{X}) + \mathcal{X},\\ 
    \label{eq:bi2}
    \mathcal{G} &= \text{Dropout}(\text{LayerNorm}(\mathcal{F}'(\mathcal{X}))),
\end{align}
where $\mathcal{G}=(\hat{\mathcal{E}}_{u_1}, ...,\hat{\mathcal{E}}_{u_k}, ...,\hat{\mathcal{E}}_{u_K})$ represents the $K$ hidden item sequences processed by forward and backward SSD blocks, and $\beta$ is a hyper-parameter to balance the contribution of backward sequence modeling.

\item \textbf{Position-wise Feed-Forward Network.} To decode the modeling of user actions in the hidden dimensions, we leverage a position-wise feed-forward network (PFFN)~\cite{liu2024mamba4rec} consisting of two dense layers:
\begin{align}
    \label{eq:pffn}
    \text{PFFN}(\mathcal{G}) = \text{Dropout}(\text{GELU}(\mathcal{G}\boldsymbol{W}^{(1)}+\boldsymbol{b}^{(1)}))\boldsymbol{W}^{(2)}+\boldsymbol{b}^{(2)},
\end{align}
where $W^{(1)}\in \mathbb{R}^{D \times 4 D}$, $W^{(2)}\in \mathbb{R}^{4D \times D}$, $b^{(1)}\in \mathbb{R}^{4D}$, $b^{(2)}\in \mathbb{R}^{D}$ are the weights and biases of the two layers, respectively. We incorporate dropout between the two dense layers in the PFFN while the activation function is set to be GELU~\cite{hendrycks2016gaussian}. PFFN can be considered a channel-mixing process that enhances the model's ability to understand and process complex patterns by mixing the information from each channel that passes through the output of bidirectional SSD blocks.
Furthermore, to enhance robustness and prevent overfitting, we incorporate residual connection, alongside dropout and layer normalization, to the PFFN. The final output of each Bi-SSD layer $\mathcal{H}$ can be calculated by:
\begin{align}
    \label{eq:pffn2}
    \mathcal{H} = \text{Dropout}(\text{LayerNorm}(\text{PFFN}(\mathcal{G}) + \mathcal{G})).
\end{align}

\item \textbf{Stacking Bi-SSD Layers.} As detailed above, user preferences throughout the entire user-item sequence can be easily captured by a single Bi-SSD layer. Furthermore, Bi-SSD layers can be stacked to allow the model to capture more complex user preferences.
However, adding more layers to the Bi-SSD increases training time due to the higher number of model parameters.
We will analyze the appropriate number of Bi-SSD layers to stack in the later experimental section.
In summary, \ourname{} refines the hidden representations of $m$-th layer as follows:
\begin{align}
    \label{eq:pffn2}
    \mathcal{H}^{m+1} &= \textbf{Bi-SSD}(\mathcal{H}^{m}, \mathcal{R}_b), \\
    \textbf{Bi-SSD}(\mathcal{G}^{m}, \mathcal{R}_b) &= \text{Dropout}(\text{LayerNorm}(\text{PFFN}(\mathcal{G}^{m}) + \mathcal{G}^{m})), \\
    \mathcal{G}^{m} &= \text{Dropout}(\text{LayerNorm}(\mathcal{F}'(\mathcal{H}^{m}))), \\
    \mathcal{F}'(\mathcal{H}^{m}) &= \mathcal{F}_{\rm forward}(\mathcal{H}^{m}) + \beta * \mathcal{F}_{\rm backward}(\mathcal{H}^{m}) + \mathcal{H}^{m},
\end{align}
where $m=0,1,..., M$, and $\mathcal{H}^0=\mathcal{X}$.

\end{itemize}

\subsubsection{\textbf{Prediction Layer}}
Given the integrated input $\mathcal{X} = \{\mathcal{E}_b, \mathcal{R}_b\}$ associated with $K$ users in the $b$-th batch, after $M$ Bi-SSD layers, we get the hidden representations of user-item sequence $\mathcal{H}^M$. Notably, in the final Bi-SSD layer, we utilize the generated representation of the last item for each user to serve as the predicted preference representation. This process can be formulated as follows:
\begin{align}
    \label{eq:predict}
    \mathcal{H}^M = [\mathbf{H}_{u_1},..., \mathbf{H}_{u_k}, ..., \mathbf{H}_{u_K}] \rightarrow \mathcal{P}_b = \{\hat{\mathbf{p}}_1, ..., \hat{\mathbf{p}}_k,..., \hat{\mathbf{p}}_K\}, 
\end{align}
where $\mathbf{H}_{u_k}$ denotes the sequence of hidden representations of user $u_k$ in $\mathcal{H}^M$. $\hat{\mathbf{p}}_k \in \mathbb{R}^D$ corresponds to the representation of the last item associated with user $u_k$ in $\mathbf{H}_{u_k}$ and acts as the predicted representation.
Using these predicted representations in a batch $\mathcal{P}_b$, \ourname{} can calculate the final output prediction scores.
In a specific batch $b$ with $K$ users, the prediction process can be written by
\begin{align}
    \label{eq:output}
    \hat{\mathcal{Y}}_b = \text{Softmax}(\mathcal{P}_b * \mathcal{Q}).
\end{align}
where $\hat{\mathcal{Y}}_b = \{\hat{\mathbf{y}}_{u_1}, ..., \hat{\mathbf{y}}_{u_K}\} \in \mathbb{R}^{1 \times K \times |\mathcal{V}|}$ represents the predicted probability distributions over the next item in the item set $\mathcal{V}$, and $\mathcal{Q}$ denotes the corresponding embeddings for items in $\mathcal{V}$.

\subsection{\textbf{Learning Objective}}
To learn the parameters of the proposed method \ourname{} for the sequential recommendation, we use a representative objective function, Cross-Entropy (CE) loss, between the predicted probability distributions and the classification labels of corresponding positive items for optimizing the model. 
Mathematically, in a batch with $K$ users, the loss function can be expressed as
\begin{align}
    \label{eq:loss}
    \mathcal{L} = - \sum^K_{k=1} \sum^D_{d=1} \mathbf{g}_{u_k,d} \log(\hat{\mathbf{y}}_{u_k,d}), 
\end{align}
where $\mathbf{g}_{u_k,d}$ and $\hat{\mathbf{y}}_{u_k,d}$ denote the $d$-th dimensions of the groundtruth distribution and predicted distribution for the user $u_k$, respectively.
Given the loss, we can update our model iteratively in a mini-batch manner.


\subsection{Discussion}
\subsubsection{\textbf{Bidirectional Learning Analysis}}
It is not straightforward and intuitive for conventional architectures to train the bidirectional model for a sequential recommendation.
For instance, jointly conditioning on both left and right context in a deep bidirectional Transformer-based model would cause information leakage, i.e., allowing each item to indirectly “see the target item”.
This could make predicting the future trivial, and the network would not learn anything useful~\cite{sun2019bert4rec}.
To avoid information leakage, the proposed \ourname{} conducts the reverse process solely on the previous items based on the property of Sequence Alignment~\cite{hwang2024hydra} in the SSD architecture.
Together with the variable-length input, our approach implements data augmentation by creating separate samples.
This means for each individual sample with a target item $v_L$, we have the forward item sequence $\{v_1, v_2, ..., v_{L-1}\}$ and the backward item sequence $\{v_{L-1}, ..., v_2, v_1\}$ as model inputs.
Although it may result in a slight decrease in training efficiency, one notable advantage of this approach is its ability to enhance the diversity of samples in each batch, unlike the Transformer-based learning process~\cite{kang2018self}, where data-augmented samples of the same user are bound to be present in the same batch.

\subsubsection{\textbf{Time Complexity Analysis}}
\label{sec:efficiency}

In general, the typical attention mechanisms require  $L^2N$ total floating-point operations (FLOPs)~\cite{mamba2}, where $L$ is the length of sequence input and $N$ is the dimensional size of hidden representations. It can be observed that the computational overhead can quadratically increase with the input size. 
The utilized SSD block involves $LN^2$ total FLOPs, scaling linearly with the sequence length $L$. 
As shown in Table~\ref{tab:complexity}, this feature in SSD enables the proposed model to attain reduced computational complexity in various recommendation scenarios characterized by prolonged user interactions, such as news recommendations and social media platforms, where the sequence length $L$ significantly exceeds the dimensional size  $N$ of hidden representations, i.e., $L >> N$. 
Furthermore, the SSD layer empowers our model to execute Transformer-like matrix multiplications, enhancing efficiency compared to a straightforward selective SSM layer that utilizes the Parallel Associative Scan ~\cite{gu2023mamba}.

\begin{table}[htbp]
  \centering
  \caption{Comparison of computational complexity between representative sequential recommendation methods, where "M.M." refers to Matrix "Multiplication".}
    \begin{tabular}{cccc}
    \toprule
          & SASRec~\cite{kang2018self} & Mamba4Rec~\cite{liu2024mamba4rec} & \textbf{\ourname{}} \\
    \midrule
    Architecture & Attention & SSM   & SSD \\
    Training FLOPs &$ L^2N$ & $LN^2$    & $LN^2$ \\
    Inference FLOPs &$ LN$ & $N^2$    & $N^2$ \\
    Memory &    $L^2$   &   $LN^2$    & $LN$ \\
    M.M. & \checkmark   & $\times$ & \checkmark \\
    \bottomrule
    \end{tabular}%
  \label{tab:complexity}%
\end{table}%

\section{Experiment}
\label{sec:Experiments}

\subsection{\textbf{Experimental Settings}}

\begin{table}[htbp]
  \centering
  \caption{Basic statistics of benchmark datasets.}
    \begin{tabular}{crrrr}
    \toprule
    \textbf{Dataset} & \multicolumn{1}{c}{\textbf{ML-1M}} & \multicolumn{1}{c}{\textbf{Beauty}} & \multicolumn{1}{c}{\textbf{Games}} & \multicolumn{1}{c}{\textbf{KuaiRand}} \\
    \midrule
    \#Users & 6,040  & 22,363 & 24,303 & 23,951 \\
    \#Items & 3,416  & 12,101 & 10,673 & 7,111 \\
    \#Interactions & 999,611 & 198,502 & 231,780 & 1,134,420 \\
    Avg.Length & 165.5 & 8.9   & 9.5   & 47.4 \\
    Max.Length & 2,314  & 389   & 880   & 809 \\
    Sparsity & 95.15\% & 99.92\% & 99.91\% & 99.33\% \\
    \bottomrule
    \end{tabular}%
  \label{tab:dataset}%
\end{table}%

\subsubsection{\textbf{Datasets}}
To showcase the efficiency and effectiveness of our proposed approach, we perform extensive experiments across four benchmark datasets: Amazon-Beauty (referred to as \textbf{Beauty}), Amazon-Video-Games (referred to as \textbf{Games}), MovieLens 1M (abbreviated as \textbf{ML1M}), and \textbf{KuaiRand}.
The initial two datasets, sourced from the Amazon e-commerce platform\footnote{\url{https://nijianmo.github.io/amazon/}}, encapsulate a wide array of user engagements with Beauty and Video Games products.
Moreover, the ML1M\footnote{\url{https://grouplens.org/datasets/movielens/1m/}} dataset comprises a compilation of movie ratings provided by users on the MovieLens platform.
Finally, the KuaiRand dataset\footnote{\url{https://kuairand.com/}}, retrieved from the recommendation logs of the video-sharing mobile application Kuaishou, includes millions of interactions involving randomly exposed items.
Notably, we filter out users and items with fewer than five interactions in each dataset.
The statistics for these four datasets are outlined in Table~\ref{tab:dataset}.
We follow the leave-one-out policy~\cite{kang2018self} for training-validation-testing partition.

\subsubsection{\textbf{Baselines}}
To demonstrate the superiority of the proposed method, four representative sequential recommendation methods and two recently developed Mamba-based methods are used as baselines, 
namely:
\begin{itemize}
    \item \textbf{Caser}~\cite{tang2018personalized}: a typical convolution-based sequential recommender that integrates Convolutional Neural Networks (CNNs) to capture sequential features in historical user-item interactions;
    \item \textbf{GRU4Rec}~\cite{hidasi2015session}: a classical recurrent-based method in sequential recommendation, which is constructed by Recurrent Neural Networks (RNNs) with Gated Recurrent Units;
    \item \textbf{BERT4Rec}~\cite{sun2019bert4rec}: a sequential recommender system based on bidirectional Transformer layers trained using the BERT-style cloze task;
    \item \textbf{SASRec}~\cite{kang2018self}: a representative approach for sequential recommendations that utilize self-attention mechanisms to enhance its modeling capabilities;
    \item \textbf{Mamba4Rec}~\cite{liu2024mamba4rec}: a pioneering model built upon the novel Structured State Space Model, namely Mamba;
    \item \textbf{SIGMA}~\cite{liu2024bidirectional}: a hybrid sequential recommendation model based on Mamba and GRU, which enhances context modeling capability by introducing Partially Flipped Mamba with a Dense Selective Gate.
\end{itemize}

\subsubsection{\textbf{Evaluation Metrics}}
To assess the recommendation results' quality, we employ three commonly utilized evaluation metrics: the top-$K$ Hit Ratio (HR@$K$), top-$K$ Normalized Discounted Cumulative Gain (NDCG@$K$), and top-$K$ Mean Reciprocal Rank (MRR@$K$)~\cite{fan2020graph,fan2022graph}.
Higher scores signify superior recommendation performance.
The average metrics for all users in the test set are provided in the performance comparison.
Furthermore, we specify the values of $K$ as 10 and 20, with 10 serving as the default for ablation experiments.
 
\subsubsection{\textbf{Hyper-parameter Settings}}
Our evaluation is implemented based on PyTorch~\footnote{\url{https://pytorch.org/}} and RecBole~\footnote{\url{https://github.com/RUCAIBox/RecBole}}.
First, to initialize the proposed Mamba-based block, we designate the model dimension as 256 and the SSM state expansion factor as 64. 
Second, we employ Sequential Grid Search to sequentially determine the two critical hyper-parameters for our model: the backward indicator $\beta$ in Eq.~\eqref{eq:bi} and the mask ratio $\rho$ in Eq.~\eqref{eq:bernoulli}.
The search ranges for both parameters are defined as \{0, 0.1, 0.2, 0.3, 0.5, 0.7, 0.9\}.
Third, the search space for the learning rate includes values of \{0.0001, 0.001, 0.01, 0.1\}, while for the batch size, we consider \{64, 256, 512, 1024, 2048\}.
The final values for these two hyper-parameters (learning rate and batch size) are set to 0.001 and 1024, respectively.
Fourth, the number of Bi-SSD cells is explored within the range of 1 to 3.
To accommodate the baselines, we set fixed sequence lengths of 200 for ML1M and 50 for other datasets. Due to the sparsity of the Amazon-beauty and Amazon-games datasets, their dropout rate is set to 0.4. In contrast, the dropout rates for MovieLens-1M and KuaiRand are set at 0.2.
Furthermore, for a fair comparison, we ensure consistency in the key hyperparameters across models while setting the other default hyperparameters for baseline methods as suggested in their respective papers. 
Finally, all experiments are conducted on a single NVIDIA 4090 GPU, and we employ the Adam~\cite{kingma2014adam} optimizer in a mini-batch manner.
The running configuration and settings are listed in Table~\ref{tab:best_hyper}. 


\begin{table}[htbp]
  \centering
  \caption{Hyperparameter Settings corresponding to different datasets}
    \begin{tabular}{c|ccccccc}
    \toprule
    Datasets & $\beta_{best}$  & $\rho_{best}$   & $L$     & Dropout & \# Bi-SSD Layers & Batch Size & Learning Rate \\
    \midrule
    ML1M  & 0.1   & 0.1   & 200   & 0.2   & 2     & 1024  & 0.001 \\
    Beauty & 0.1   & 0.2   & 50    & 0.4   & 1     & 1024  & 0.001 \\
    Games & 0.1   & 0.1   & 50    & 0.4   & 1     & 1024  & 0.001 \\
    KuaiRand & 0.1   & 0.2   & 50    & 0.2   & 2     & 1024  & 0.001 \\
    Range & [0,1) & [0,1) & -     & -     & \{1,2,3\} & \{4, 256, 512, 1024, 2048\} & \{0.0001, 0.001, 0.01, 0.1\} \\
    \bottomrule
    \end{tabular}%
  \label{tab:best_hyper}%
\end{table}%

\subsection{\textbf{Recommendation Performance Comparison}}
We first compare the recommendation performance between the proposed model and all baselines over the four benchmark datasets.
Table~\ref{tab:result} showcases the overall evaluation results.
From the table, we can make the following observations.

\begin{itemize} 
    \item Our proposed \ourname{} consistently outperforms all the baselines across all datasets regarding NDCG@10\&20 and MRR@10\&20.
    On average, \ourname{} significantly exceeds the strongest baselines by 13.80\% on NDCG@10 and 16.14\% on MRR@10 in the ML1M dataset.
    Such improvement demonstrates the effectiveness of our proposed method and the great potential of exploring the Structured State Space Duality property in sequential recommendations. 
    
    \item While inferior to the proposed model, Mamba4Rec, in most cases, outperforms the conventional deep learning methods, namely the CNN-based model Caser, RNN-based model GRU4Rec, and Attention-based models BERT4Rec and SASRec.
    This highlights the efficacy of Mamba in capturing user behavior dynamics for recommendation purposes.

    \item Additionally, it is evident that the benefits of the mamba-based models diminish in the datasets characterized by shorter average sequences, i.e., Beauty and Games. 
    Conversely, their performance excels in datasets like ML1M and KuaiRand, where the average sequence lengths extend to 165.5 and 47.4, respectively.
    This disparity in performance can be attributed to Mamba's exceptional capacity to capture long-range dependencies.
\end{itemize}

\begin{table}[htbp]
  \centering
  \caption{Sequential recommendation performance on four different datasets, where Imp.* denotes the relative performance improvement in comparison to the specified baseline marked by *. The best and second-best results are bold and underlined, respectively.}
    \begin{tabular}{cc|ccccccc|c}
    \toprule
    \textbf{Dataset} & \textbf{Metric} & \textbf{Caser} & \textbf{GRU4Rec} & \textbf{BERT4Rec} & \textbf{SASRec*} & \textbf{Mamba4Rec} & \textbf{SIGMA} & \textbf{\ourname{}} & \textbf{Imp.*} \\
    \midrule
    \multirow{6}[2]{*}{\textbf{ML1M}}
    & \textbf{NDCG@10} & 0.1714 & 0.1655 & 0.1846 & 0.1718 &  \underline{0.1847}  &  0.1777 & \textbf{0.1955} & 13.80\% \\
    & \textbf{NDCG@20} & 0.1948 & 0.1906 & 0.2082 & 0.1963 &  \underline{0.2094}  &  0.2026 & \textbf{0.2225} & 13.35\% \\
    & \textbf{MRR@10} & 0.1354 & 0.1276 & 0.1441 & 0.1320 &  \underline{0.1456}  &  0.1361 & \textbf{0.1533} & 16.14\% \\
    & \textbf{MRR@20} & 0.1418 & 0.1344 & 0.1506 & 0.1386 &  \underline{0.1523}  &  0.1431 & \textbf{0.1607} & 15.95\% \\
    & \textbf{HR@10} & 0.2892 & 0.2901 &  \underline{0.3171}  & 0.3023 & 0.3124 & 0.3136 & \textbf{0.3334} & 10.29\% \\
    & \textbf{HR@20} & 0.3820 & 0.3897 & 0.4103 & 0.4000 & 0.4103 & \underline{0.4121} & \textbf{0.4406} & 10.15\% \\
    \midrule
    \multirow{6}[2]{*}{\textbf{Beauty}} 
    & \textbf{NDCG@10} & 0.0294 & 0.0307 & 0.0219 & 0.0395 &  \underline{0.0442}  & 0.0376 & \textbf{0.0470} & 18.99\% \\
    & \textbf{NDCG@20} & 0.0355 & 0.0386 & 0.0279 & 0.0480 &  \underline{0.0501} & 0.0455 & \textbf{0.0535} & 11.46\% \\
    & \textbf{MRR@10} & 0.0222 & 0.0221 & 0.0159 & 0.0261 &  \underline{0.0361} & 0.0284 & \textbf{0.0383} & 46.74\% \\
    & \textbf{MRR@20} & 0.0239 & 0.0242 & 0.0175 & 0.0284 &  \underline{0.0376} & 0.0305 & \textbf{0.0400} & 40.85\% \\
    & \textbf{HR@10} & 0.0531 & 0.0593 & 0.0419 &  \textbf{0.0829} & 0.0709 & 0.0678 & \underline{0.0756} & -8.81\% \\
    & \textbf{HR@20} & 0.0774 & 0.0907 & 0.0656 &  \textbf{0.1168} & 0.0942 & 0.0994 & \underline{0.1016} & -13.01\% \\
    \midrule
    \multirow{6}[2]{*}{\textbf{Games}} 
    & \textbf{NDCG@10} & 0.0460 & 0.0520 & 0.0372 & 0.0531 &  \underline{0.0605} & 0.0592 & \textbf{0.0611} & 15.07\% \\
    & \textbf{NDCG@20} & 0.0587 & 0.0654 & 0.0482 & 0.0689 & 0.0738 & \underline{0.0746} & \textbf{0.0747} & 8.41\% \\
    & \textbf{MRR@10} & 0.0330 & 0.0373 & 0.0263 & 0.0330 &  \underline{0.0461} & 0.043 & \textbf{0.0464} & 40.61\% \\
    & \textbf{MRR@20} & 0.0364 & 0.0409 & 0.0293 & 0.0373 &  \underline{0.0497} & 0.0472 & \textbf{0.0500} & 34.05\% \\
    & \textbf{HR@10} & 0.0891 & 0.1011 & 0.0735 & 0.1189  & 0.1081 & \textbf{0.1342} & \underline{0.1099} & -7.57\% \\
    & \textbf{HR@20} & 0.1396 & 0.1543 & 0.1170 & 0.1817 & 0.1610 & \textbf{0.2002} & \underline{0.1635} & -10.02\% \\
    \midrule
    \multirow{6}[2]{*}{\textbf{KuaiRand}} 
    & \textbf{NDCG@10} & 0.0545 & 0.0563 & 0.0534 &  0.0567 & 0.0558 & \underline{0.0572} &\textbf{0.0593} & 4.59\% \\
    & \textbf{NDCG@20} & 0.0692 & 0.0722 & 0.0683 &  \underline{0.0733} & 0.0710 & 0.0727 & \textbf{0.0757} & 3.27\% \\
    & \textbf{MRR@10} & 0.0414 & 0.0426 & 0.0404 & 0.0426 & 0.0427 & \underline{0.0433} & \textbf{0.0448} & 5.16\% \\
    & \textbf{MRR@20} & 0.0453 & 0.0470 & 0.0444 & 0.0471 & 0.0468 & \underline{0.0475} &\textbf{0.0493} & 4.67\% \\
    & \textbf{HR@10} & 0.0982 & 0.1017 & 0.0968 &  \underline{0.1040} & 0.0994 & 0.1036 & \textbf{0.1075} & 3.37\% \\
    & \textbf{HR@20} & 0.1571 & 0.1653 & 0.1563 &  \underline{0.1705} & 0.1601 & 0.1655 &\textbf{0.1731} & 1.52\% \\
    \bottomrule
    \end{tabular}%
  \label{tab:result}%
\end{table}%

\begin{table}[htbp]
  \centering
  \caption{Comparison of the computational efficiency on the ML1M dataset, which involves measuring \textbf{the training time per epoch for all test users in seconds (s)}, where Imp.* denotes the relative computational speed improvement in comparison to the specified baseline marked by *; and SSD4Rec- represents the proposed model without the backward SSD block. The results indicate that SSD4Rec accelerates the training efficiency of the typical Attention-based and Mamba-based methods (i.e., SASRec and Mamba4Rec) by an average of 84.18\% and 88.53\%, respectively.}
    \begin{tabular}{cc|ccccc}
    \toprule
    \multicolumn{2}{c|}{\textbf{ML1M}} & \multicolumn{5}{c}{\textbf{Training time}} \\
    \textbf{FIX} & \textbf{$N/L$} & \textbf{SASRec}* & \textbf{Mamba4Rec} & \textbf{SSD4Rec-} & \textbf{SSD4Rec} & \textbf{Imp.}* \\
    \midrule
    \multirow{4}[2]{*}{$L=200$} 
    & 32 & 88.16s & 32.92s & 25.85s & 43.06s & 104.71\% \\
    & 64 & 91.93s & 52.06s & 26.91s & 43.86s & 109.57\% \\
    & 128 & 102.48s & 90.76s & 27.92s & 49.13s & 108.58\% \\
    & 256 & 134.29s & 172.71s & 45.56s & 79.71s & 68.49\% \\
    \midrule
    \multirow{4}[2]{*}{$N=256$} 
    & 50 & 38.53s & 51.17s & 23.79s & 39.30s & -1.96\% \\
    & 100 & 53.08s & 66.59s & 28.42s & 47.29s & 12.24\% \\
    & 200 & 134.29s & 172.71s & 45.56s & 79.71s & 68.49\% \\
    & 400 & 349.94s & 489.20s & 87.60s & 115.38s & 203.27\% \\
    \bottomrule
    \end{tabular}
  \label{tab:efficiency}%
\end{table}%

\begin{table}[htbp]
  \centering
  \caption{Comparison of the computational efficiency on the ML1M dataset, which involves measuring \textbf{the inference time for all test users in millisecond (ms)}, where Imp.* denotes the relative computational speed improvement in comparison to the specified baseline marked by *; and SSD4Rec- represents the proposed model without the backward SSD block. The results indicate that SSD4Rec surpasses the inference efficiency of the typical Attention-based and Mamba-based methods (i.e., SASRec and Mamba4Rec) by an average of 72.35\% and 53.63\%, respectively.}
    \begin{tabular}{cc|ccccc}
    \toprule
    \multicolumn{2}{c|}{\textbf{ML1M}} & \multicolumn{5}{c}{\textbf{Inference time}} \\
    \textbf{FIX} & \textbf{$N/L$} & \textbf{SASRec}* & \textbf{Mamba4Rec} & \textbf{SSD4Rec-} & \textbf{SSD4Rec} & \textbf{Imp.}* \\
    \midrule
    \multirow{4}[2]{*}{$L=200$} 
    & 32 & 116ms & 40ms & 34ms & 58ms & 100.00\% \\
    & 64 & 126ms & 58ms & 38ms & 53ms & 137.74\% \\
    & 128 & 139ms & 93ms & 41ms & 66ms & 110.61\% \\
    & 256 & 176ms & 180ms & 68ms & 122ms & 44.26\% \\
    \midrule
    \multirow{4}[2]{*}{$N=256$} 
    & 50 & 33ms & 82ms & 29ms & 47ms & -29.79\% \\
    & 100 & 71ms & 100ms & 45ms & 79ms & -10.13\% \\
    & 200 & 176ms & 180ms & 68ms & 122ms & 44.26\% \\
    & 400 & 496ms & 552ms & 96ms & 176ms & 181.81\% \\
    \bottomrule
    \end{tabular}
  \label{tab:efficiency}%
\end{table}%

\subsection{Efficiency Analysis}
In this subsection, we will mainly analyze the training and inference efficiency of \ourname{} compared with two representative sequential recommendation methods, i.e., SASRec and Mamba4Rec. For fairness, it should be noted that the number of network layers for SSD4Rec is limited to 1. This corresponds with the single-layer structures of SASRec and Mamba4Rec.
Due to limitations in GPU memory, we set both the training and evaluation batch sizes to 512 to prevent possible out-of-memory issues when N=256 and L=400.
For a comprehensive evaluation across diverse sequence length $L$ and token dimension $N$, we fix one of the two parameters while varying the other. 
We also incorporate \textbf{SSD4Rec-} as a baseline in our efficiency analysis for training and inference time, which simply removes the backward SSD block structure from our processed SSD4Rec model.
The evaluation results on the ML1M dataset are presented in Table~\ref{tab:efficiency}.
We can see that the two Mamba-based models, i.e., Mamba4Rec and \ourname{} are much more efficient compared to the Attention-based SASRec model when $N<L=200$, which in line with the discussion in Section~\ref{sec:efficiency}, i.e., Mamba enjoys linear scaling in sequence length. 
When presented with inputs of substantial dimensions, such as $N=256$, Mamba4Rec exhibits lower efficiency compared to SASRec, even when $L=400>N$, showcasing the constraints of Mamba-1 in managing high-dimensional inputs~\cite{mamba2}.
In contrast, SSD4Rec achieves superior computational efficiency to SASRec and Mamba4Rec in most cases.
Specifically, \ourname{} only showcases slightly inferior training and inference speeds compared to the baselines when  $N$ is significantly larger than $L$ (i.e., $N=256$ and $L=50/100$)
However, it demonstrates a clear advantage as $L$ increases due to its linear computational complexity with respect to sequence length, as discussed in Table~\ref{tab:complexity}.
Moreover, this superior performance can be attributed to the utilization of the advancing hardware-aware computation algorithm (i.e., the block-based matrix multiplication~\cite{mamba2}) in our SSD-based model.
For instance, when given a lengthy interaction list, i.e., $N=32$ and $L=200$, the proposed model is 104\% and 100\% faster than the Attention-based method SASRec in the training and inference phases, respectively.
This substantial acceleration underscores a notable enhancement in computational efficiency.
Even when configured with high-dimensional input sequences, i.e., $N=256>L=200$, our model experiences a speedup of exceeding 40\% in comparison to SASRec for both the training and inference tasks. 
Moreover, the \ourname{}- model without backward SSD block demonstrates superior computational efficiency compared to SASRec across all scenarios, including cases where $L$ is significantly smaller than $N$. These two observations highlight the substantial performance gains achieved through the advanced hardware-aware algorithm integrated within the SSD block.

\begin{table}[htbp]
  \centering
  \caption{Ablation studies conducted by eliminating the effects of each key component.
  }
    \begin{tabular}{lcccccc}
    \toprule
    \multirow{2}[2]{*}{\textbf{Component}} & \multicolumn{3}{c}{\textbf{ML1M}} & \multicolumn{3}{c}{\textbf{KuaiRand}} \\
          & NDCG@10 & MRR@10 & HR@10 & NDCG@10 & MRR@10 & HR@10 \\
    \midrule
    Full & 0.1955 & 0.1533 & 0.3334 & 0.0593 & 0.0448 & 0.1075 \\
    w/o Bidirection & 0.1921 & 0.1521 & 0.3222 & 0.0585 & 0.0446 & 0.1049 \\
    w/o mask & 0.1897 & 0.1488 & 0.3235 & 0.0581 & 0.0446 & 0.1034 \\
    w/o Varlen & 0.1891 & 0.1484 & 0.3222 & 0.0572 & 0.0436 & 0.1025 \\
    w/o Register & 0.1920 & 0.1509 & 0.3262 & 0.0565 & 0.0429 & 0.1022 \\
    w PE  & 0.1901 & 0.1496 & 0.3225 & 0.0585 & 0.0447 & 0.1049 \\
    \midrule
    1 Layer & 0.1930 & 0.1523 & 0.3258 & 0.0584 & 0.0442 & 0.1055 \\
    3 Layers & 0.1926 & 0.1505 & 0.3303 & 0.0557 & 0.0416 & 0.1020 \\
    \bottomrule
    \end{tabular}%
  \label{tab:ablation}%
\end{table}%

\subsection{Ablation Study}
In order to assess the effectiveness of the proposed key components,
we conducted ablation experiments on the ML1M and KuaiRand datasets, where the influence of each component was eliminated separately as follows:

\begin{itemize}
    \item w/o Bidirection: Deactivate the backward SSD layers to eliminate backward information of user interactions.
    \item w/o mask: Remove the mask operation on the model input, i.e., all inputs remain in their original state.
    \item w/o Varlen: Deploy a fixed window on the model input through padding and truncation. In contrast to the proposed variable-length input, there is a tendency to lose information from the long-tail user interaction histories.
    \item w/o Register: Remove the user-aware segment registers from the model input. By doing so, the model will consider the integrated sequence containing diverse interactions from different users as a successive item list.
    \item w PE: Incorporate learnable position embeddings into the item embedding sequences.
\end{itemize}

From the ablation results in Table~\ref{tab:ablation}, we can see that each component in our approach contributes to the overall performance since eliminating any of them would result in performance degradation.
Moreover, the results demonstrate that the integration of learnable positional embeddings does not improve performance due to the recurrent nature of the Mamba block.
Finally, stacking Mamba layers yields performance improvements on the ML1M dataset with larger, long-sequence user interactions, although it would lead to a linear increase in the model size. 
To this end, as shown in Table~\ref{tab:best_hyper}, we have designated a 2-layer \ourname{} model for the ML1M and KuaiRand datasets and a 1-layer \ourname{} model for the Beauty and Games datasets, respectively.

\subsection{Hyper-parameter Analysis}
In \ourname{}, we introduce three critical hyper-parameters, namely the weighted indicator of backward processing $\beta$, the mask ratio indicator $\rho$ and the max length of item sequences $L$.
Their value sensitivities are evaluated in this section to facilitate the future application of our proposed model.

\subsubsection{\textbf{Effect of Backward Indicator $\beta$}}
\begin{figure*}[htbp]
\centering
{\subfigure[ML1M - NDCG@10]
{\includegraphics[width=0.24\linewidth]{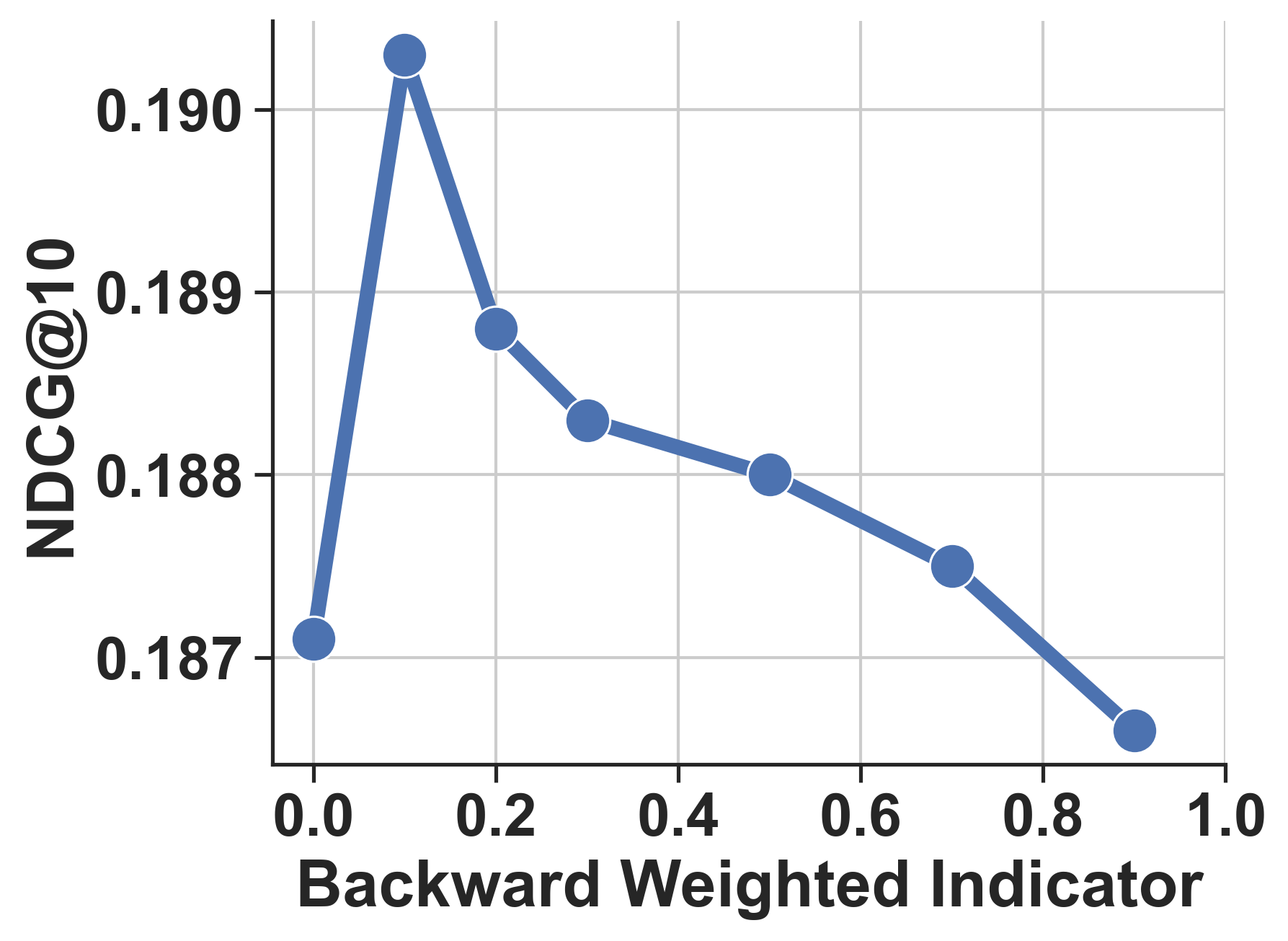}}}
{\subfigure[Beauty - NDCG@10]
{\includegraphics[width=0.24\linewidth]{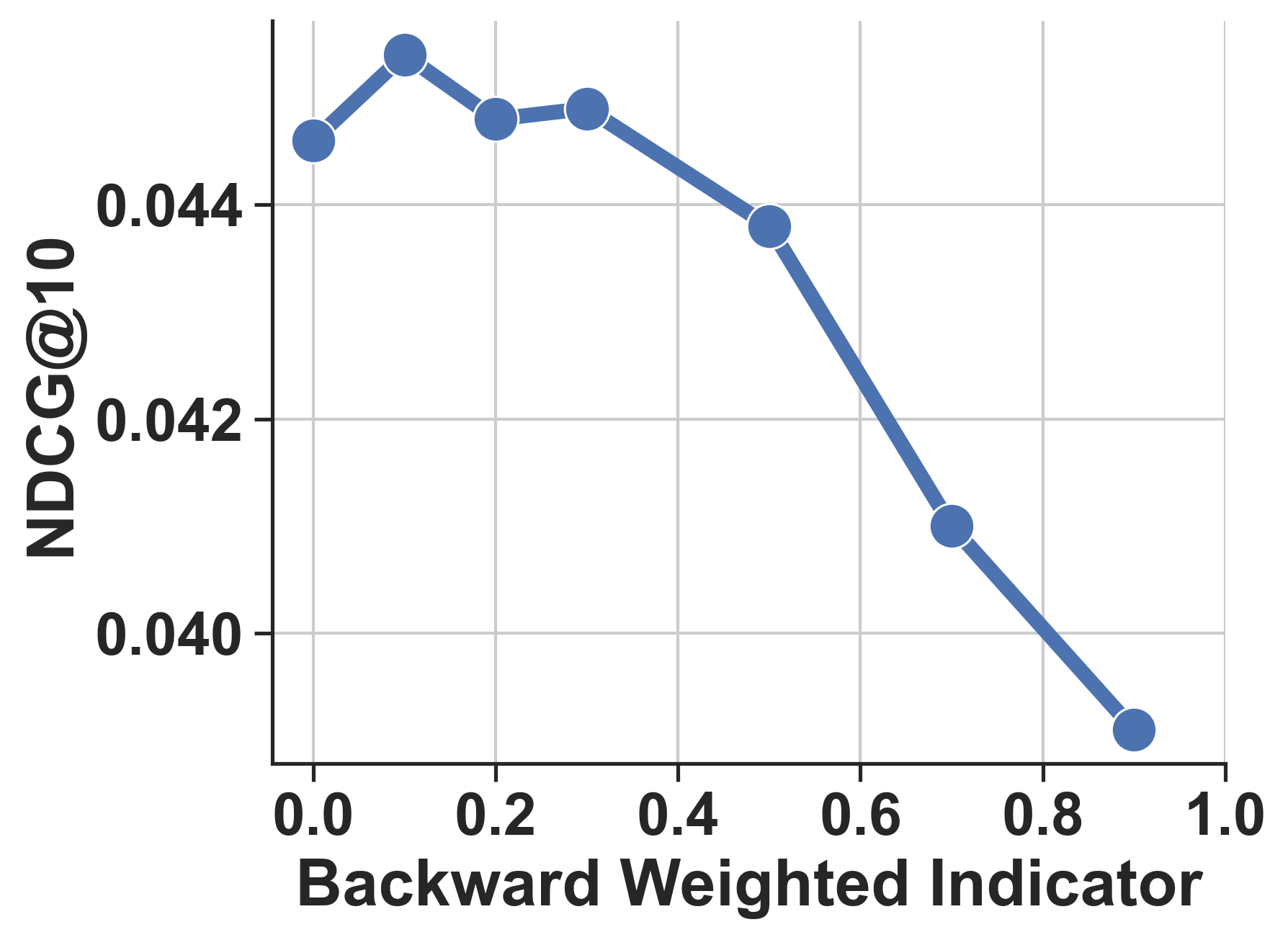}}}
{\subfigure[Games - NDCG@10]
{\includegraphics[width=0.24\linewidth]{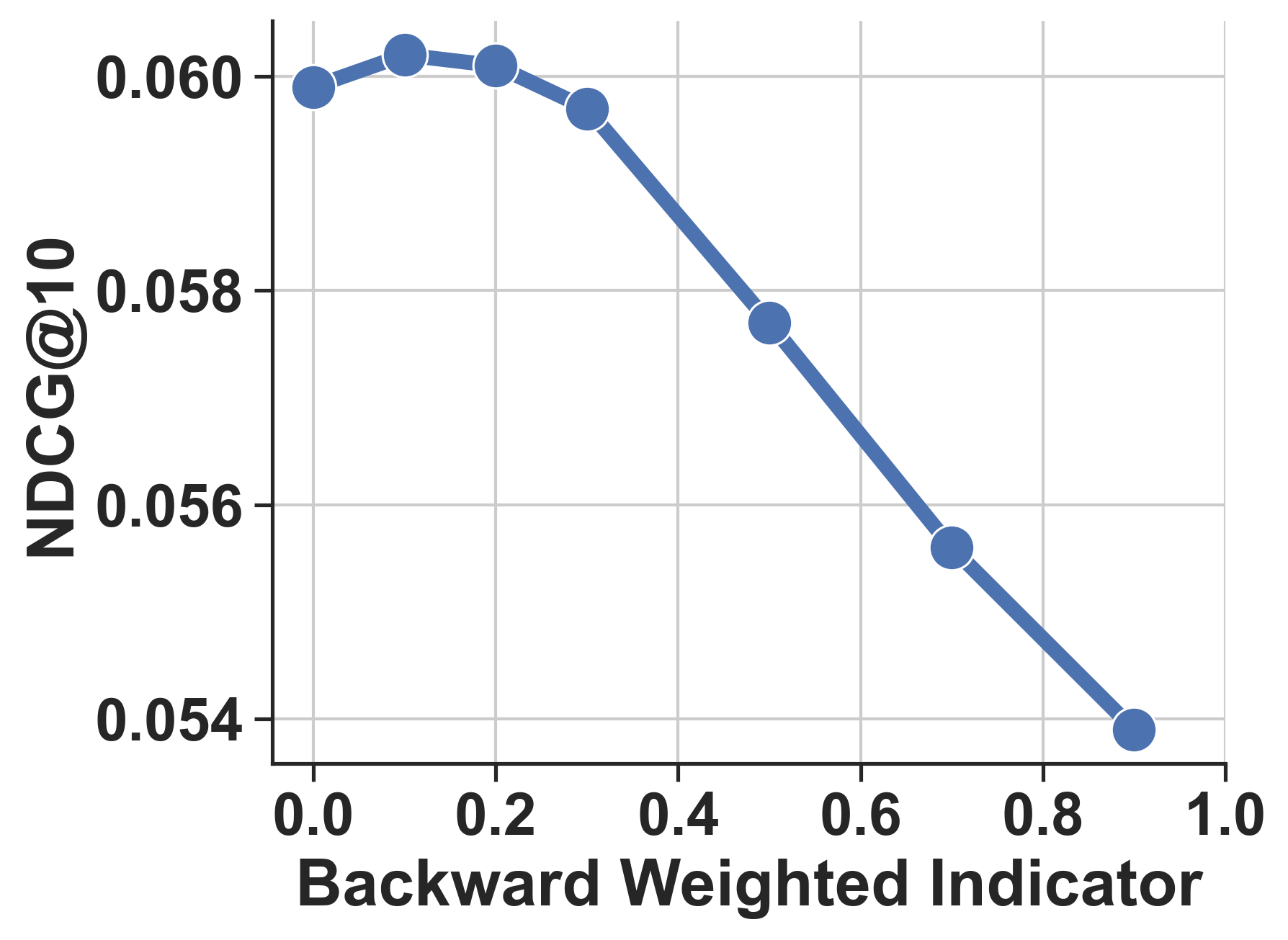}}}
{\subfigure[KuaiRand - NDCG@10]
{\includegraphics[width=0.24\linewidth]{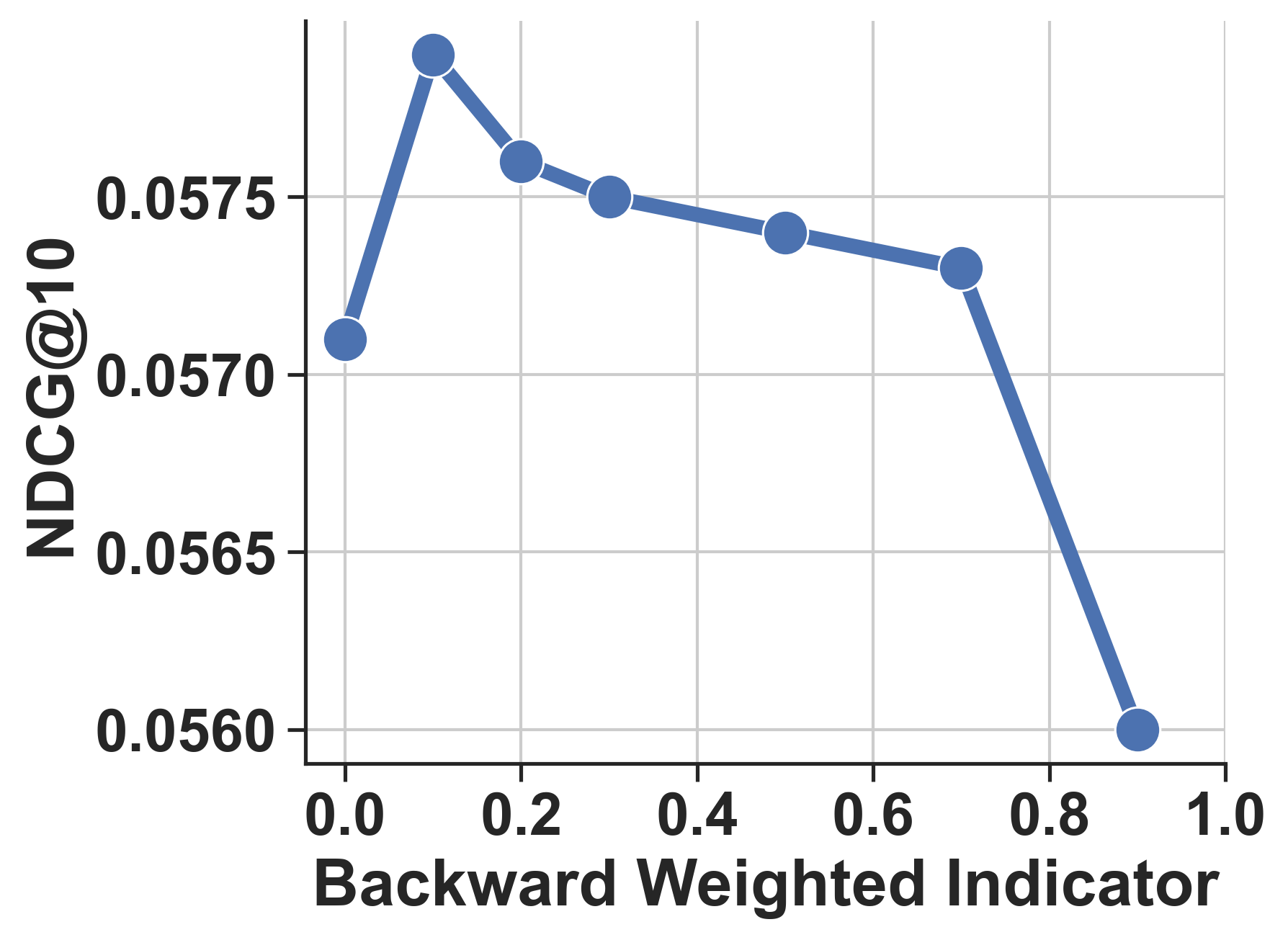}}}

\vspace{1em} 

{\subfigure[ML1M - MRR@10]
{\includegraphics[width=0.24\linewidth]{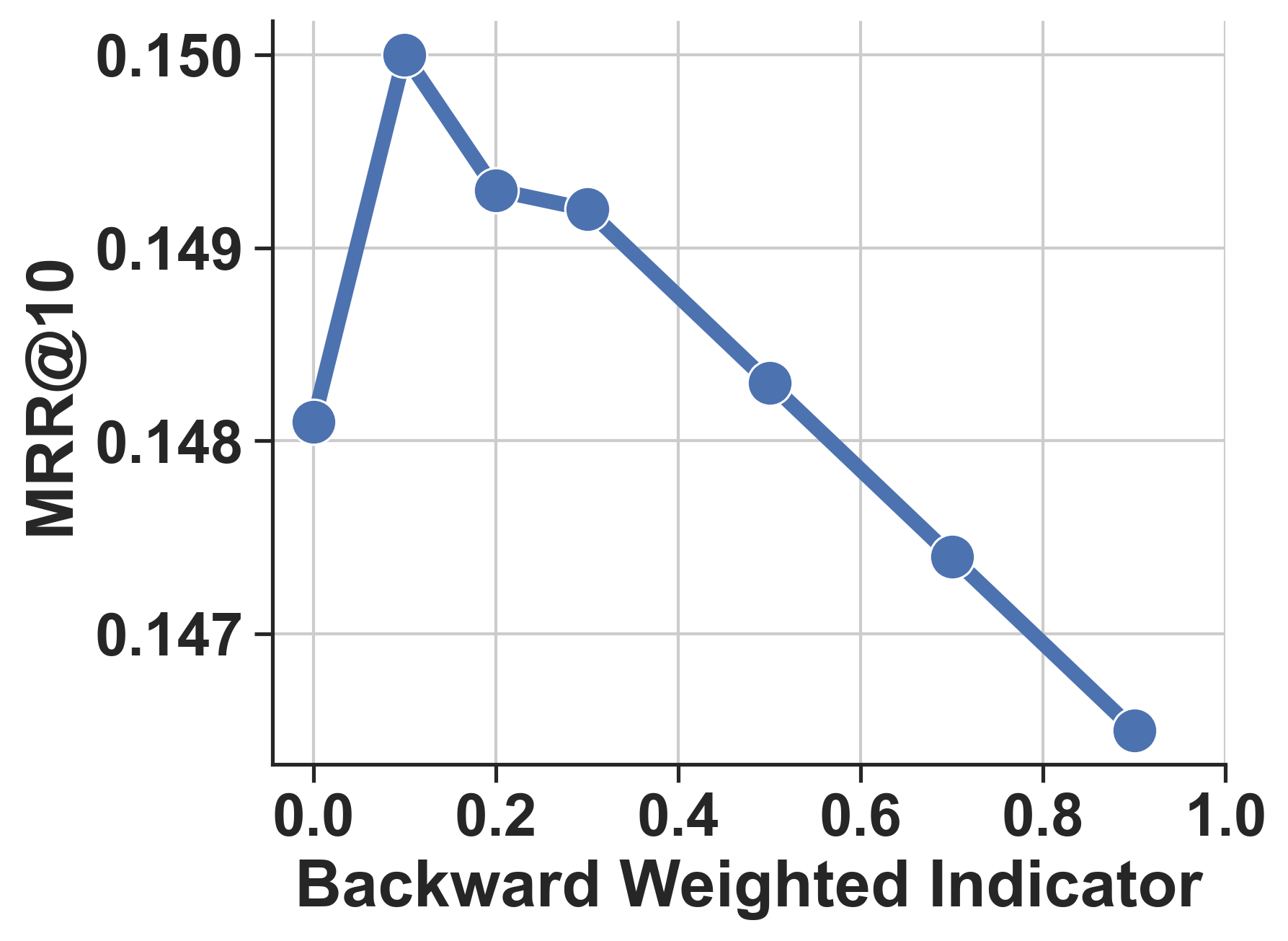}}}
{\subfigure[Beauty - MRR@10]
{\includegraphics[width=0.24\linewidth]{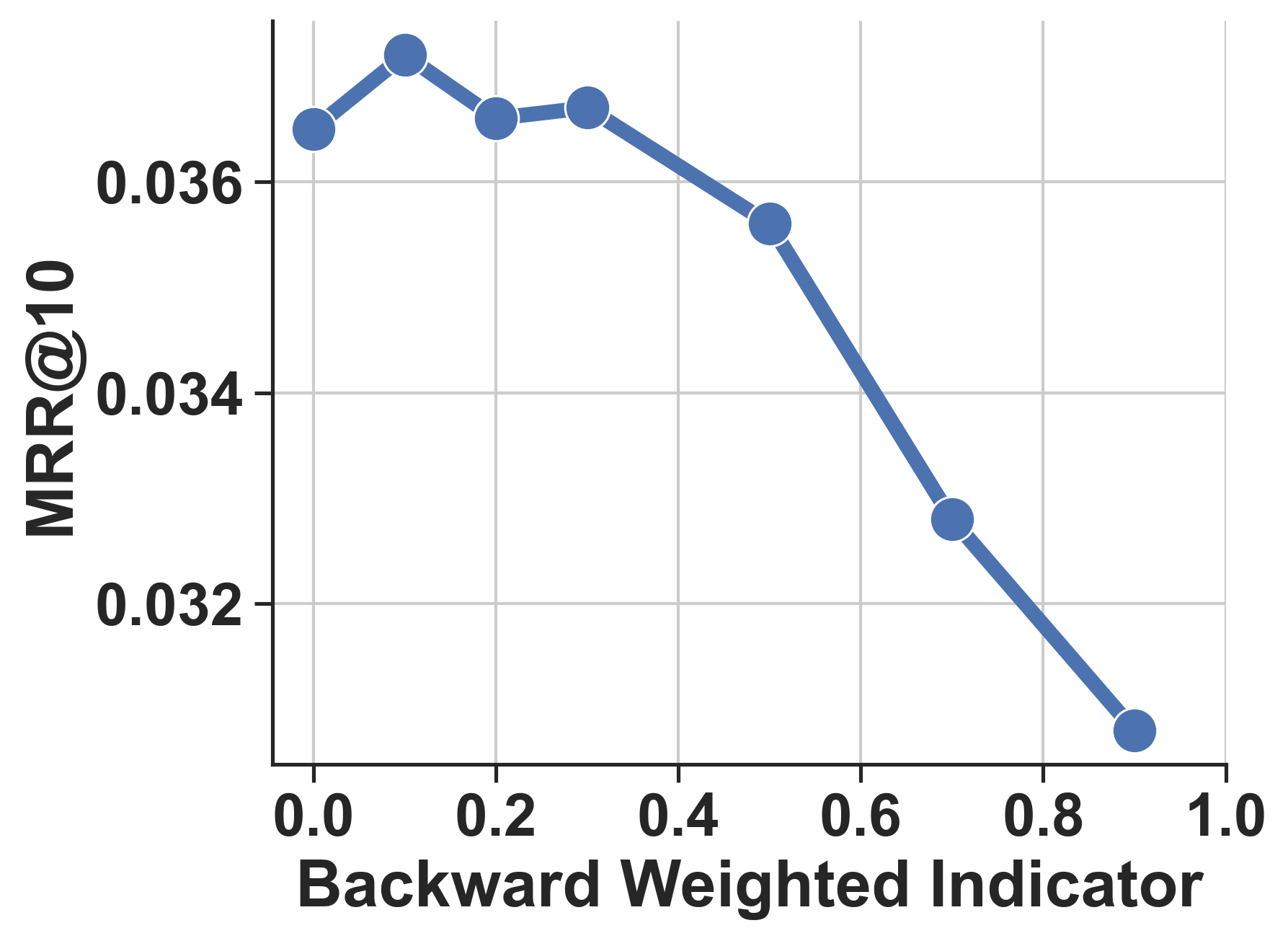}}}
{\subfigure[Games - MRR@10]
{\includegraphics[width=0.24\linewidth]{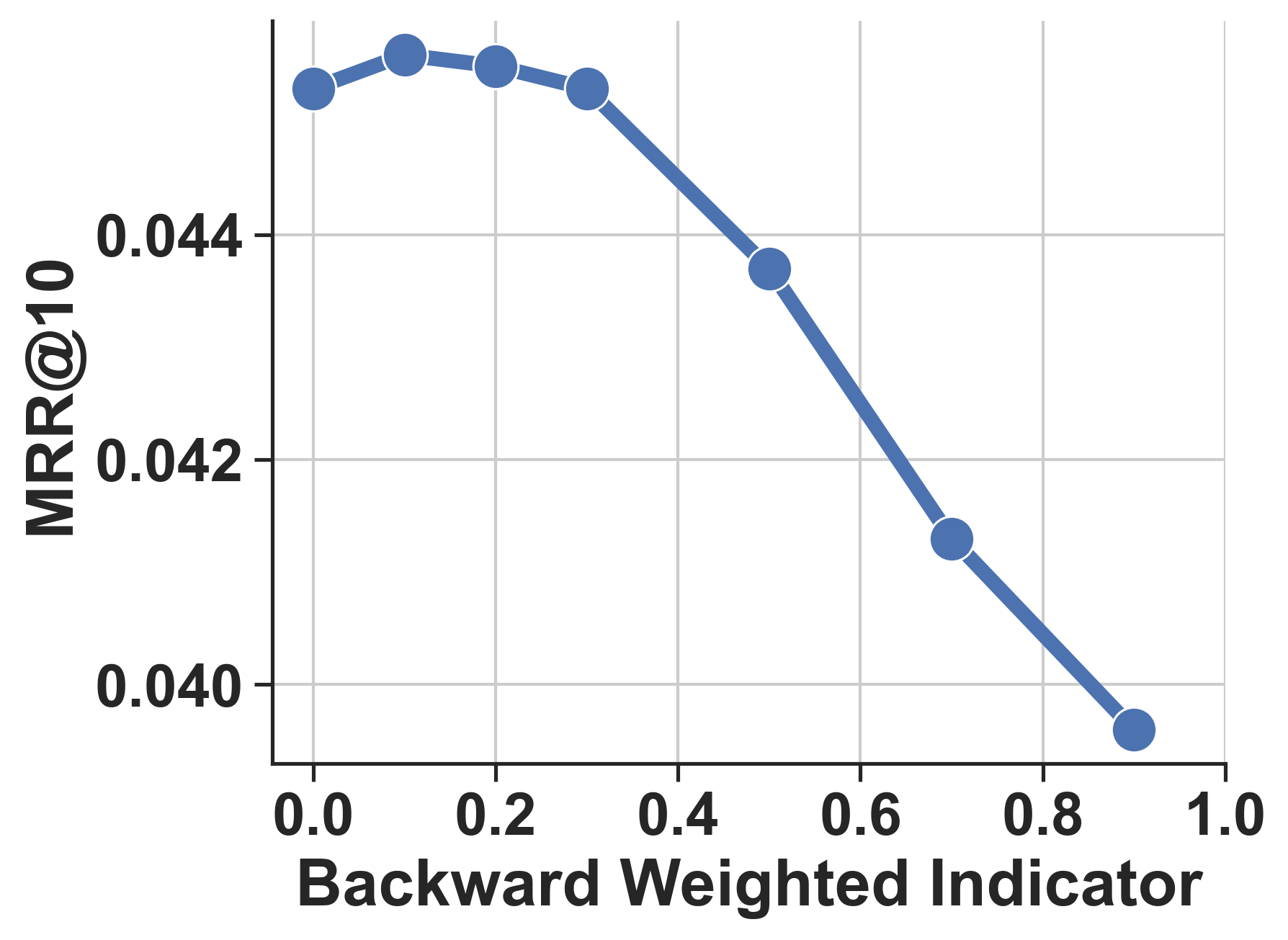}}}
{\subfigure[KuaiRand - MRR@10]
{\includegraphics[width=0.24\linewidth]{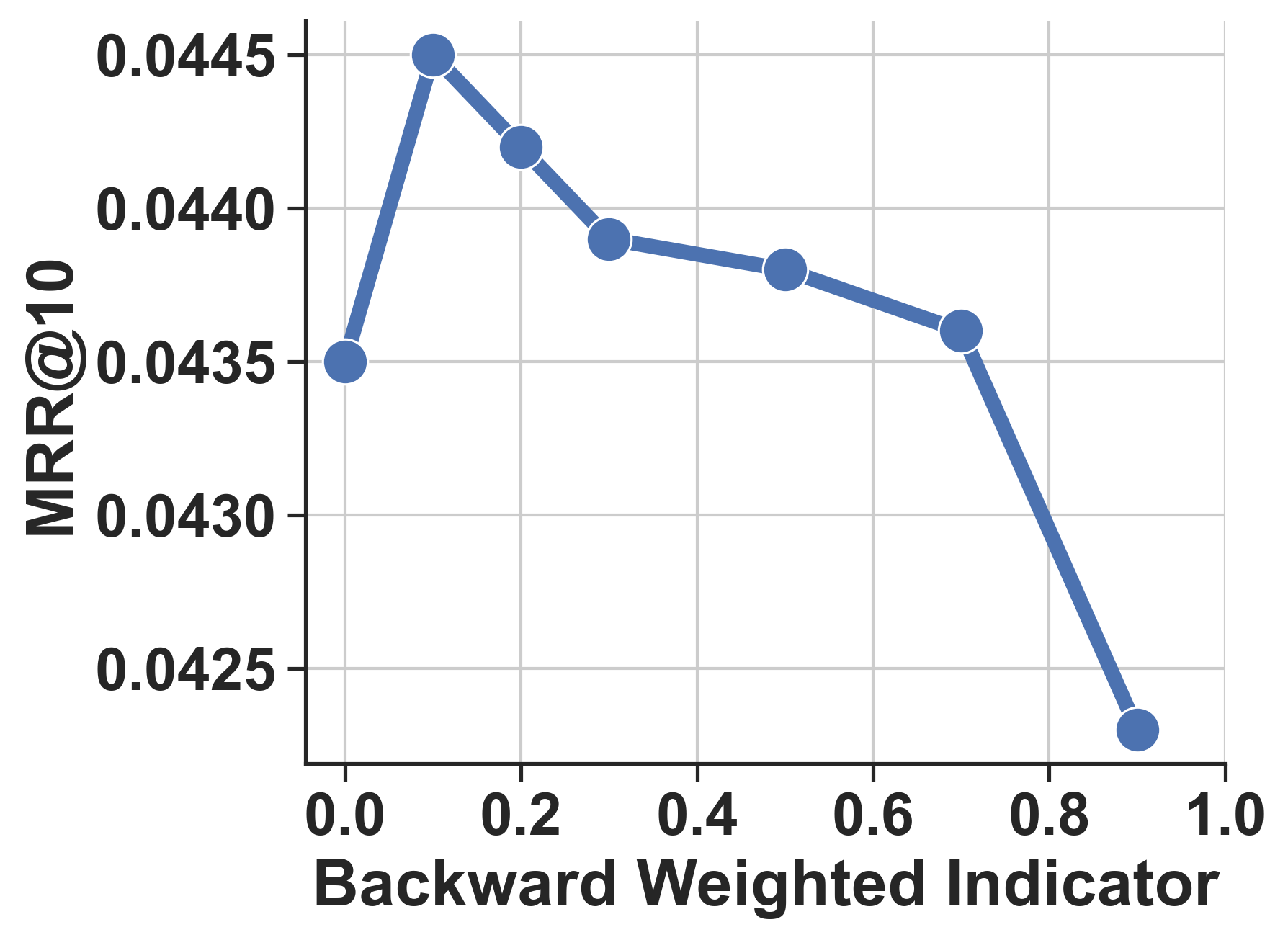}}}

\vspace{1em} 

{\subfigure[ML1M - HR@10]
{\includegraphics[width=0.24\linewidth]{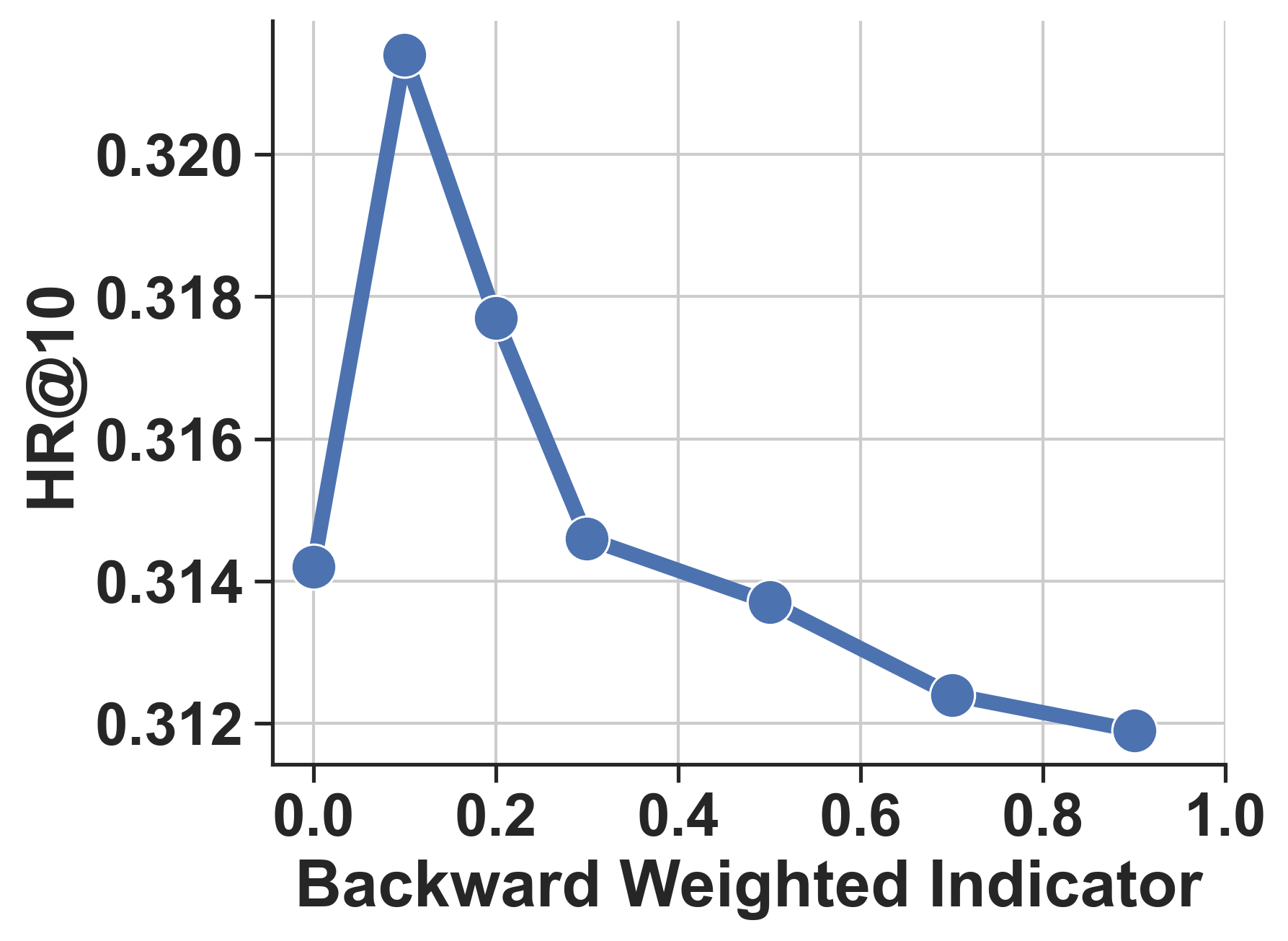}}}
{\subfigure[Beauty - HR@10]
{\includegraphics[width=0.24\linewidth]{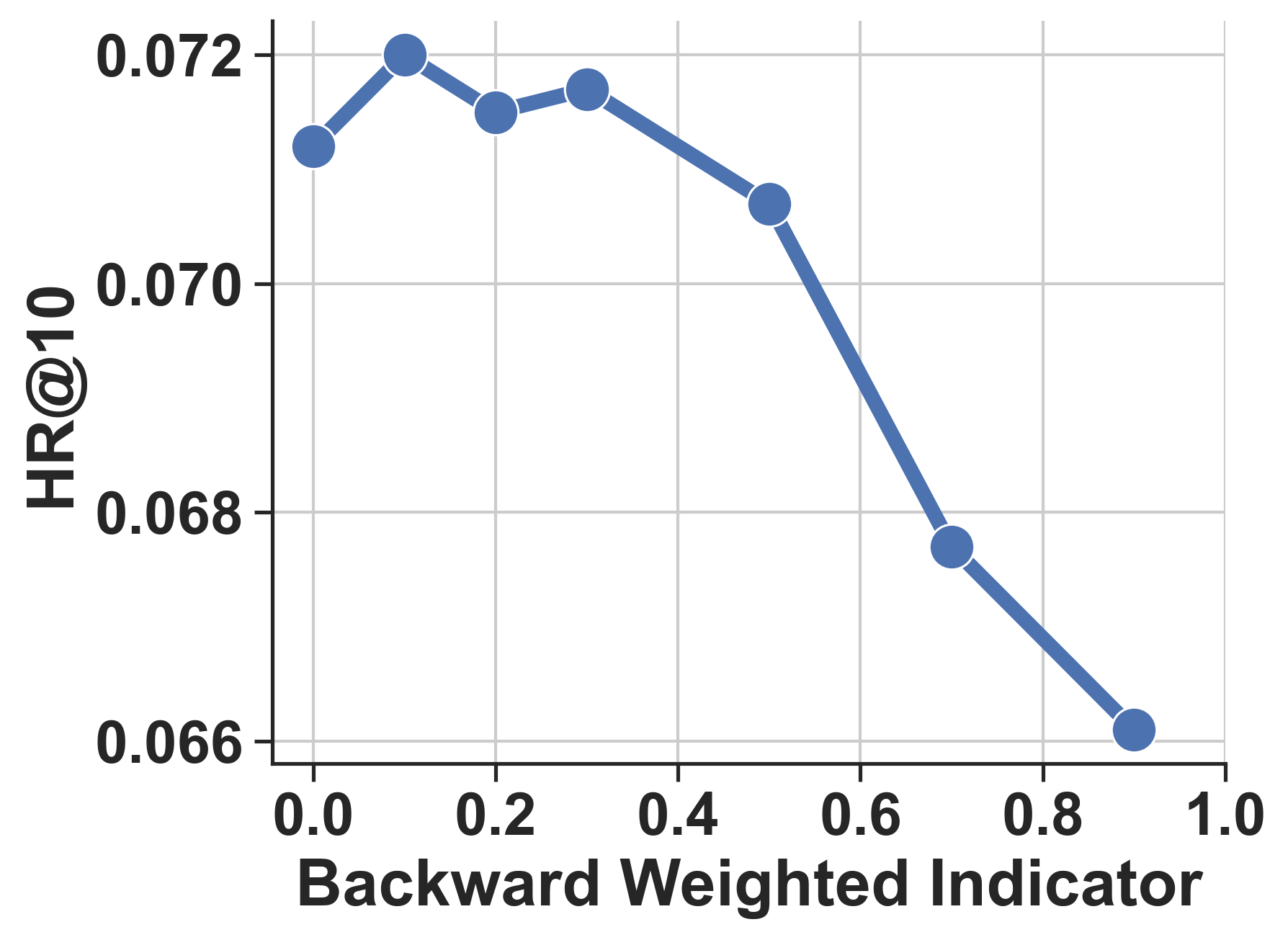}}}
{\subfigure[Games - HR@10]
{\includegraphics[width=0.24\linewidth]{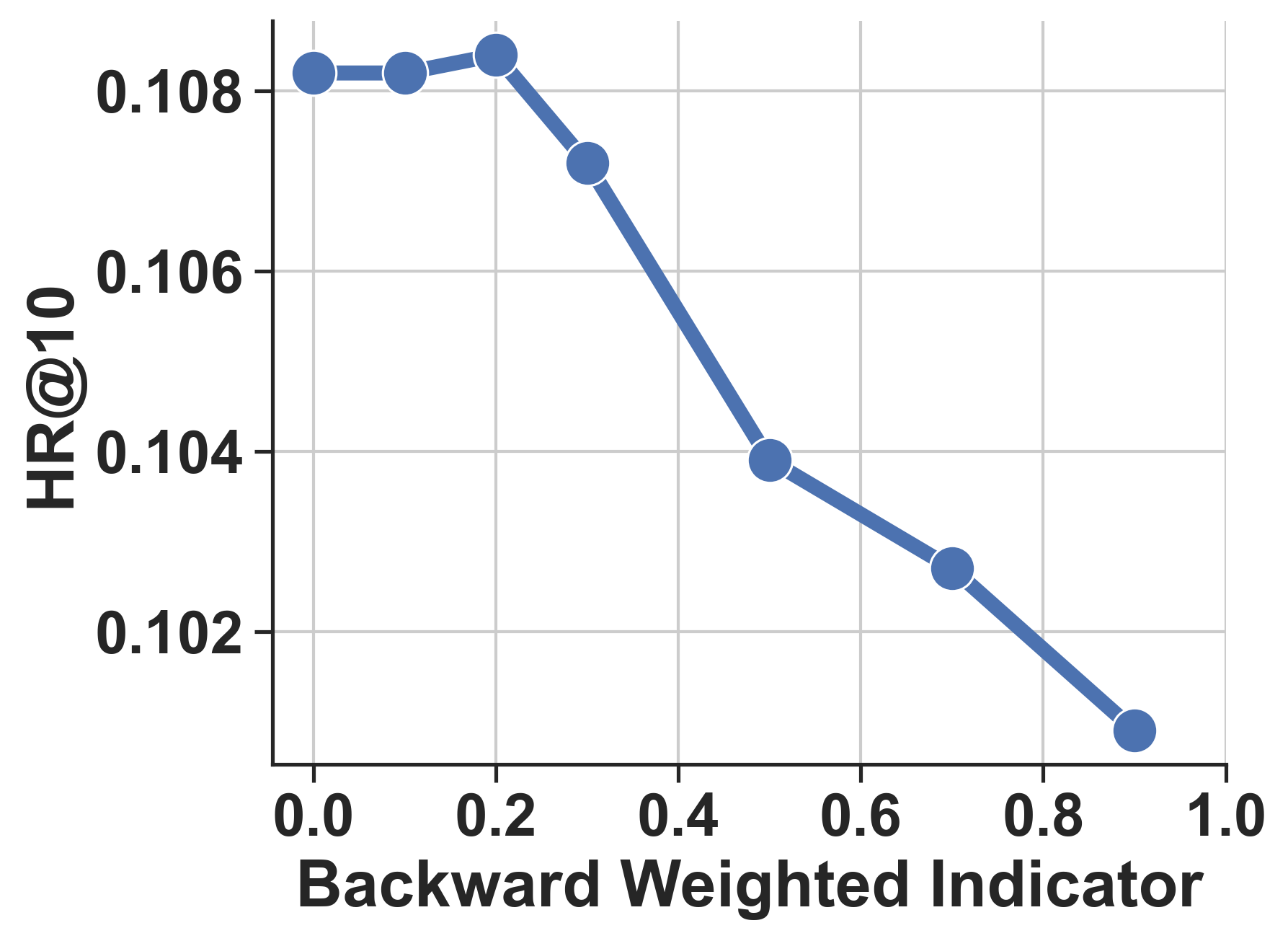}}}
{\subfigure[KuaiRand - HR@10]
{\includegraphics[width=0.24\linewidth]{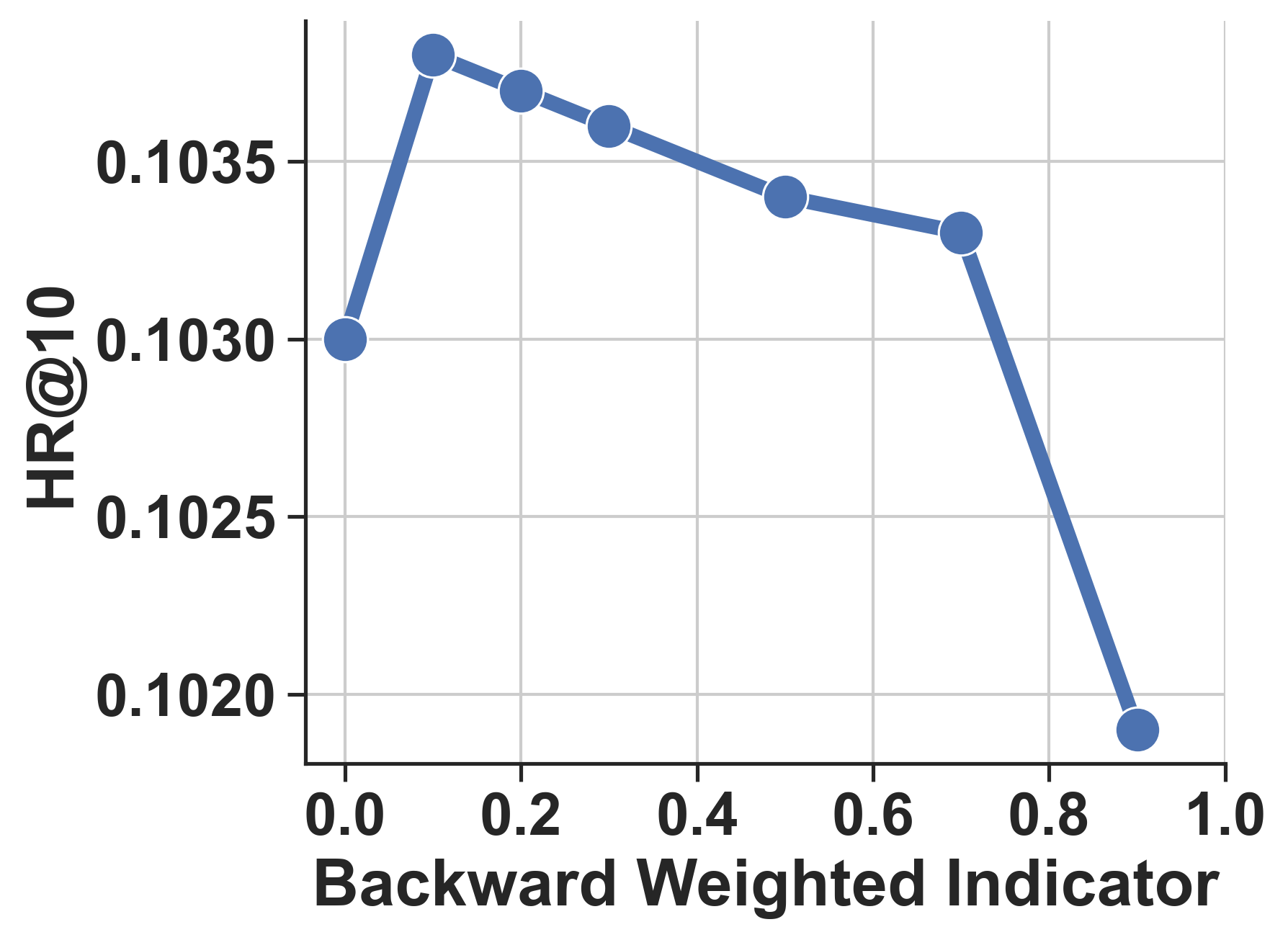}}}
\caption{The effect of backward weighted indicator $\beta$ under NDCG@10, MRR@10, HR@10, separately.}
\label{fig:beta_ndcg}
\end{figure*}

This hyper-parameter determines the degree to which backward information can propagate through the SSD layer.
Figure~\ref{fig:beta_ndcg} shows the corresponding performance change of \ourname{} on NDCG@10, MRR@10, HR@10, separately.
We can find that introducing a small ratio of backward knowledge brings performance improvements.
In most cases, the recommendation performance of our proposed method
improves when $\beta< 0.3$.
The experimental results also reveal that the recommendation
performance degrades when the backward indicator $\beta\geq 0.5$, suggesting excessive backward information should be avoided. In most cases, the model demonstrates optimal performance when $\beta$ is set to 0.1. Therefore, we designate $\beta = 0.1$ as the default value for our proposed method, \ourname{}.


\subsubsection{\textbf{Effect of Mask Ratio indicator $\rho$}}
\begin{figure*}[htbp]
\centering
{\subfigure[ML1M - NDCG@10]
{\includegraphics[width=0.24\linewidth]{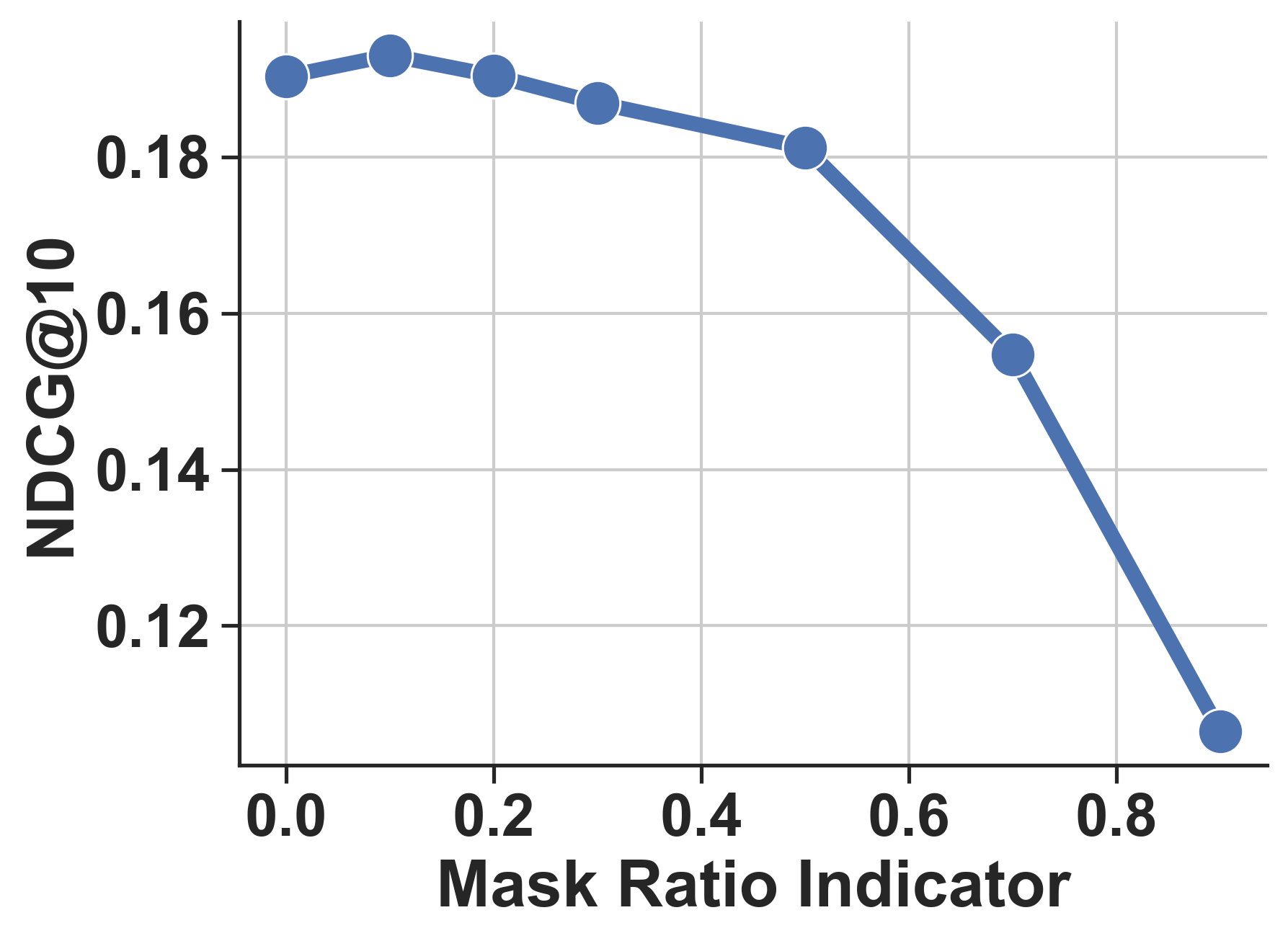}}}
{\subfigure[Beauty - NDCG@10]
{\includegraphics[width=0.24\linewidth]{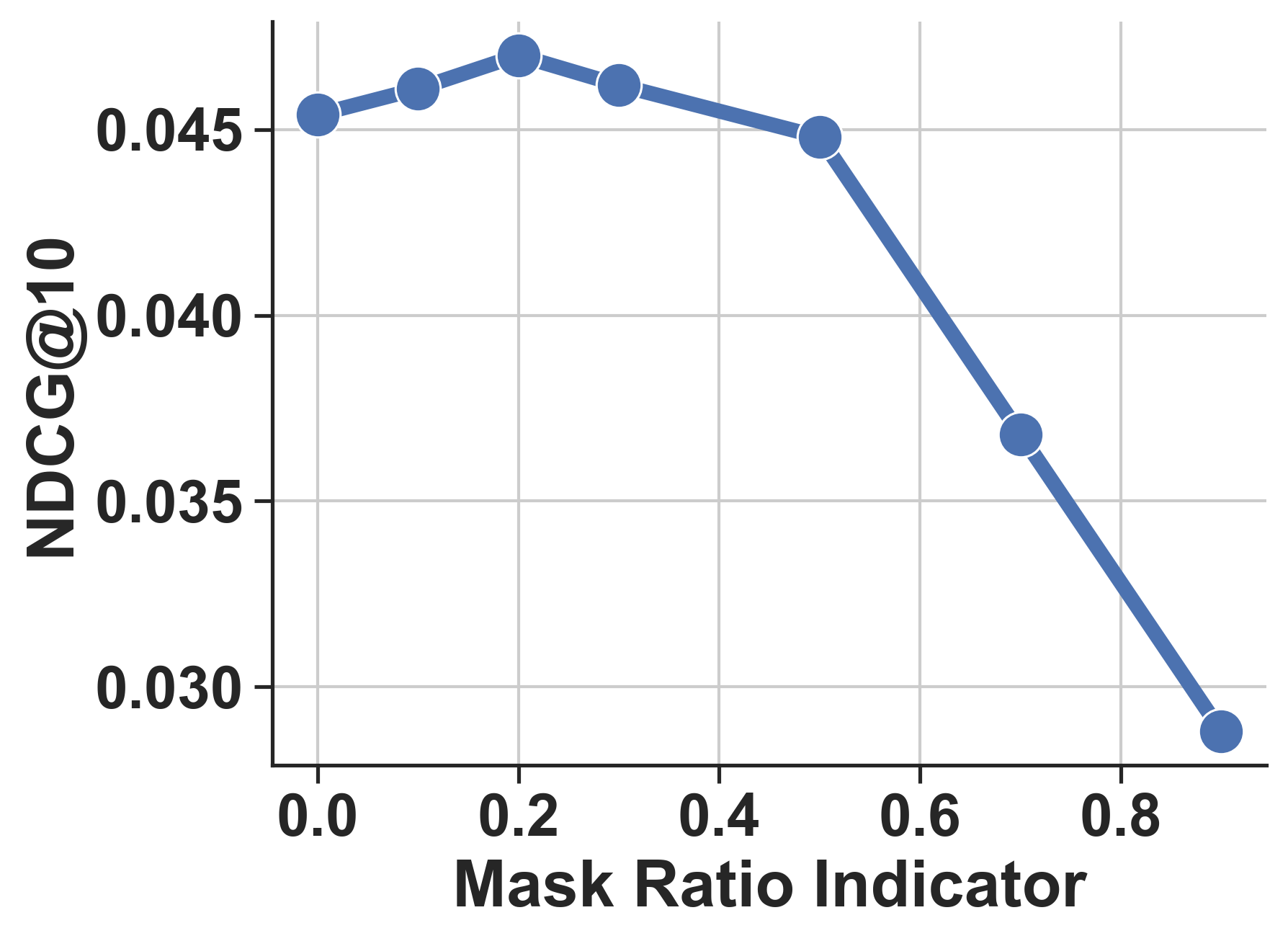}}}
{\subfigure[Games - NDCG@10]
{\includegraphics[width=0.24\linewidth]{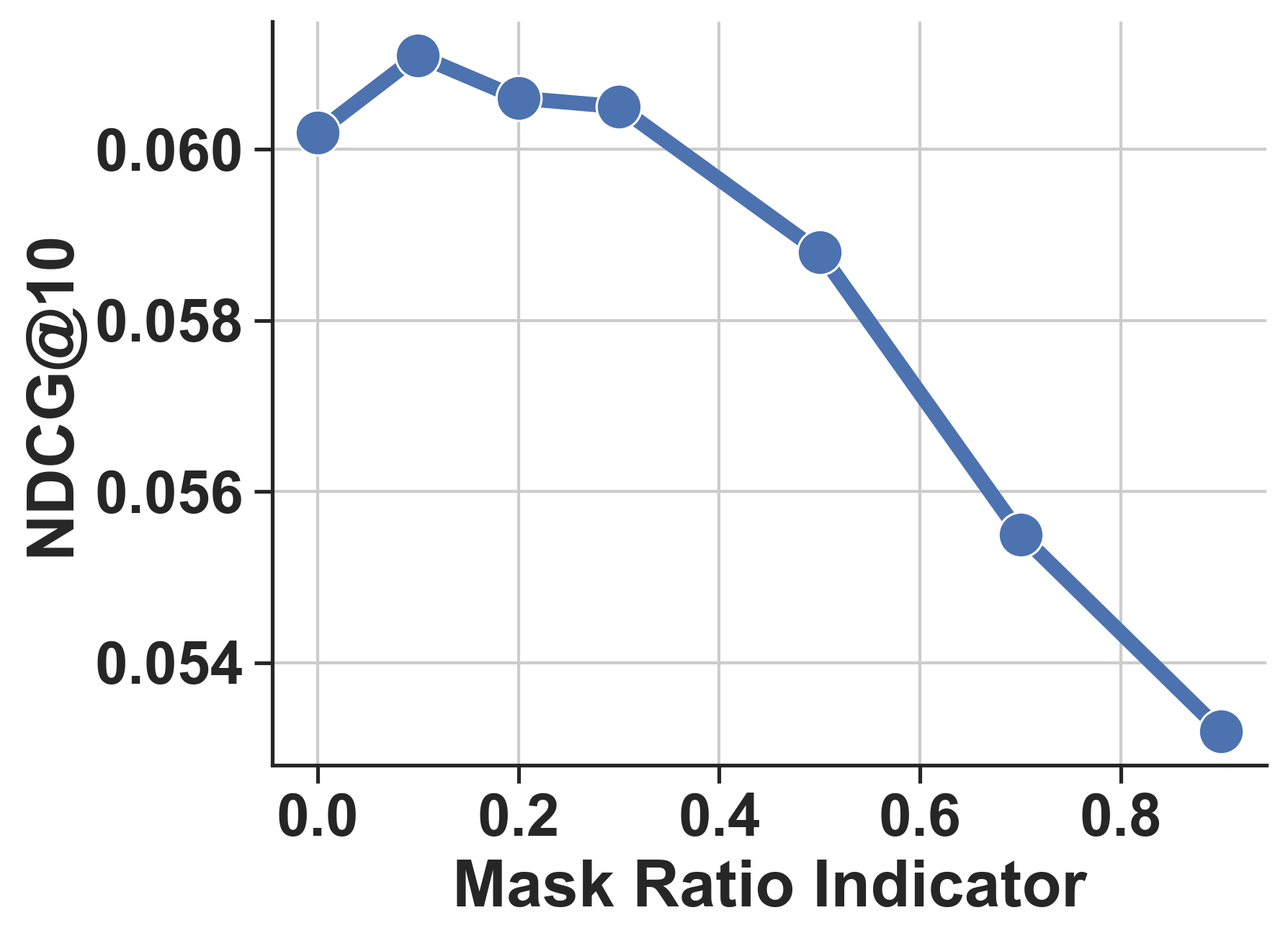}}}
{\subfigure[KuaiRand - NDCG@10]
{\includegraphics[width=0.24\linewidth]{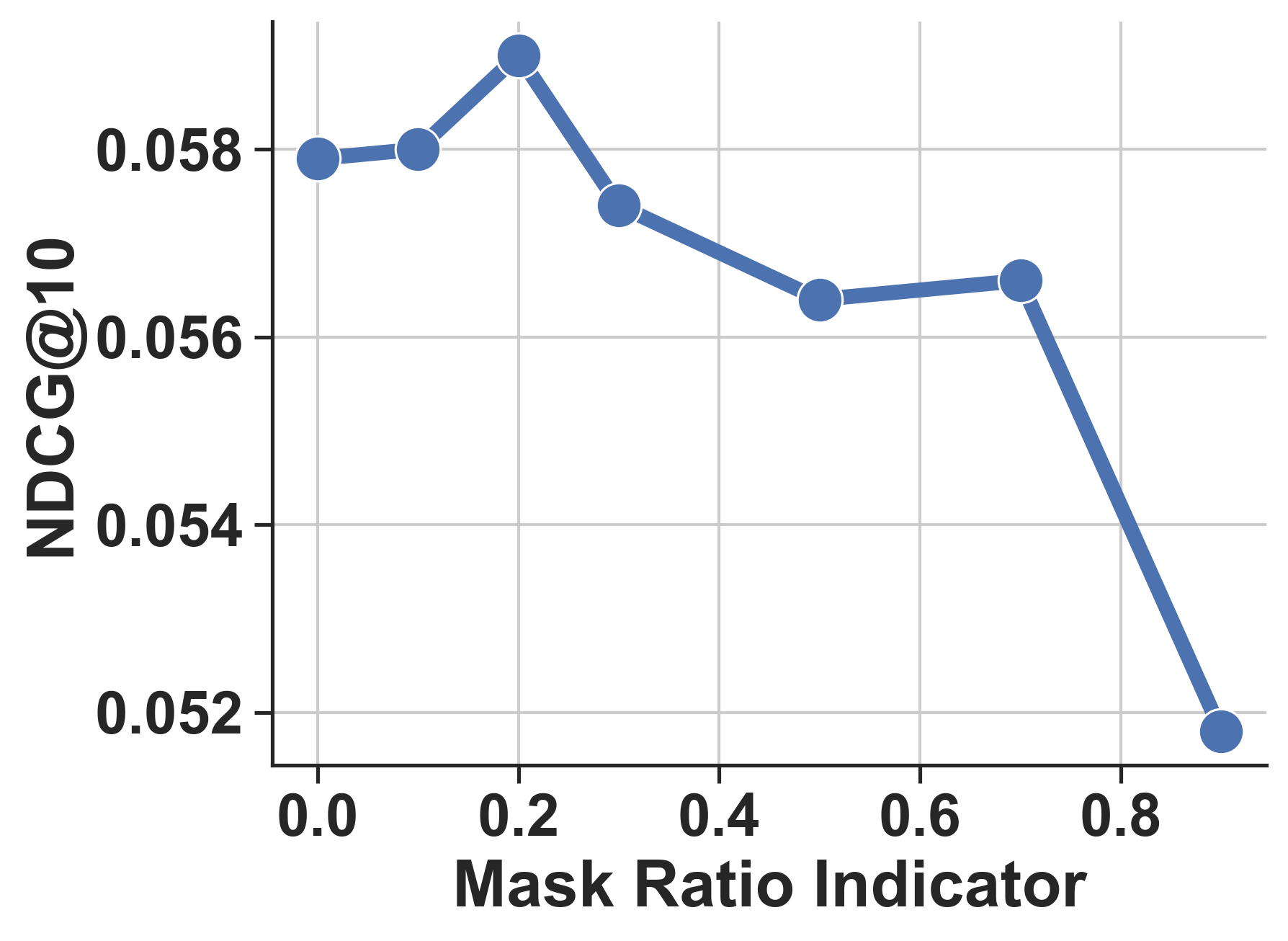}}}

\vspace{1em} 

{\subfigure[ML1M - MRR@10]
{\includegraphics[width=0.24\linewidth]{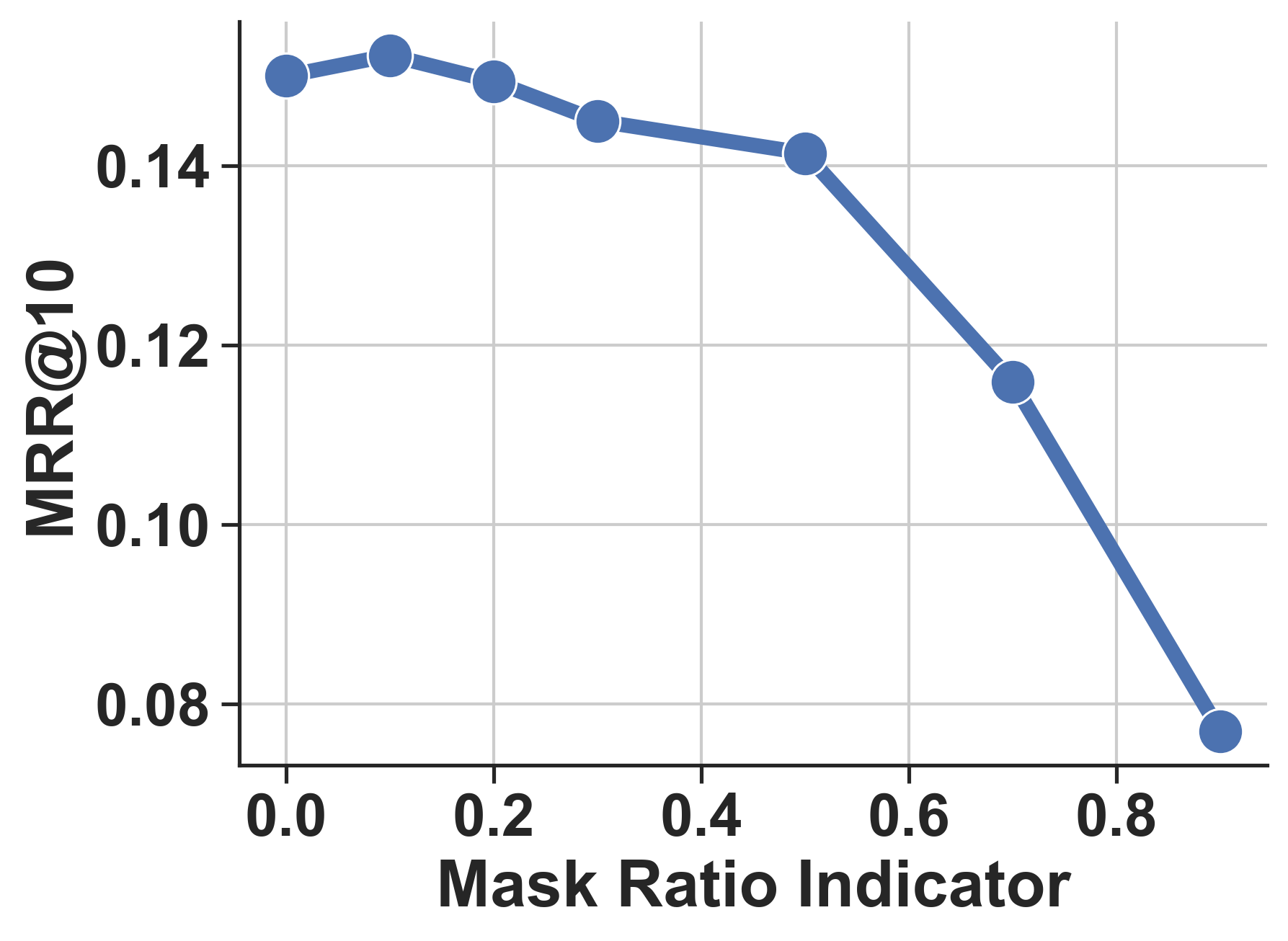}}}
{\subfigure[Beauty - MRR@10]
{\includegraphics[width=0.24\linewidth]{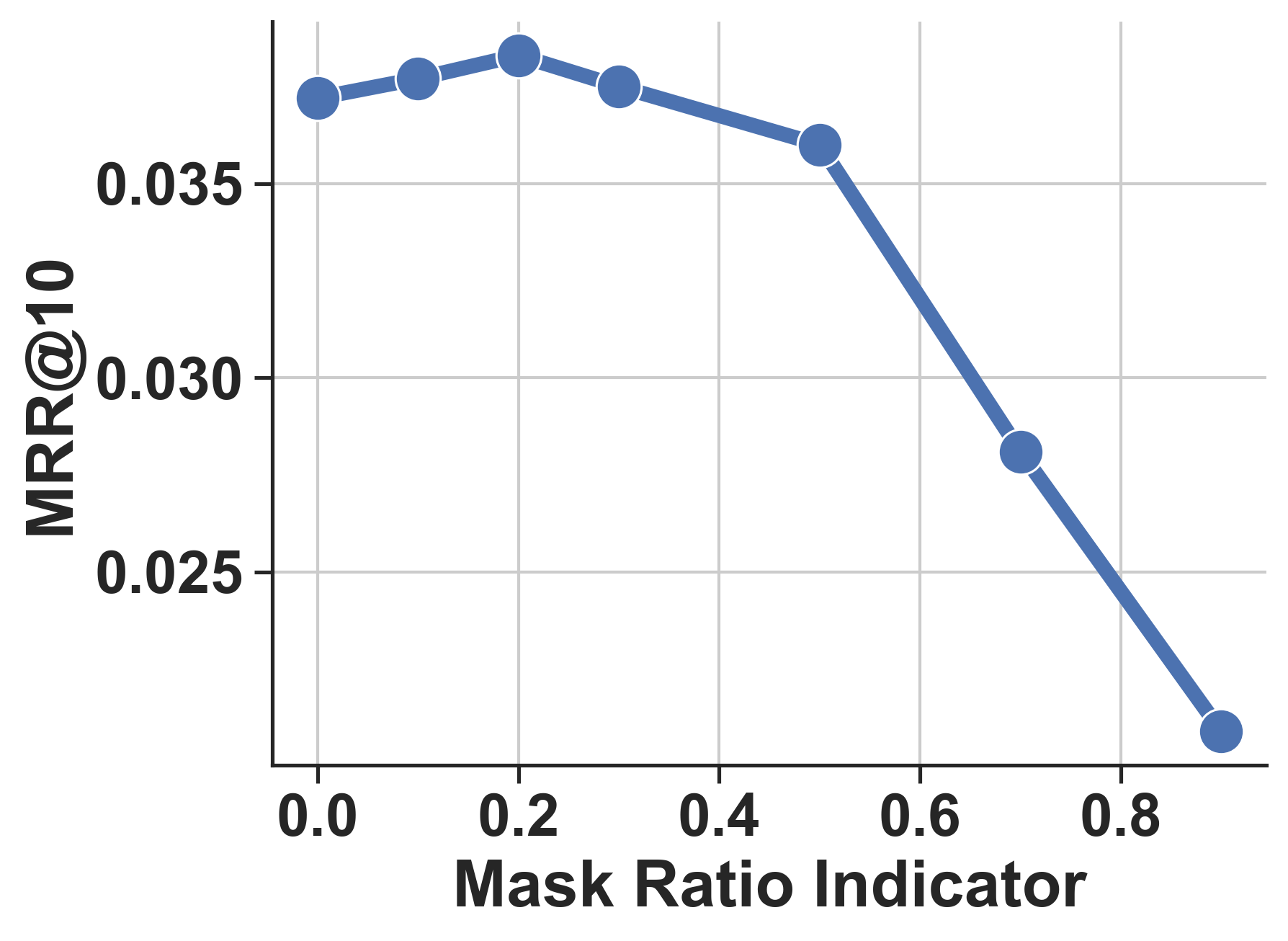}}}
{\subfigure[Games - MRR@10]
{\includegraphics[width=0.24\linewidth]{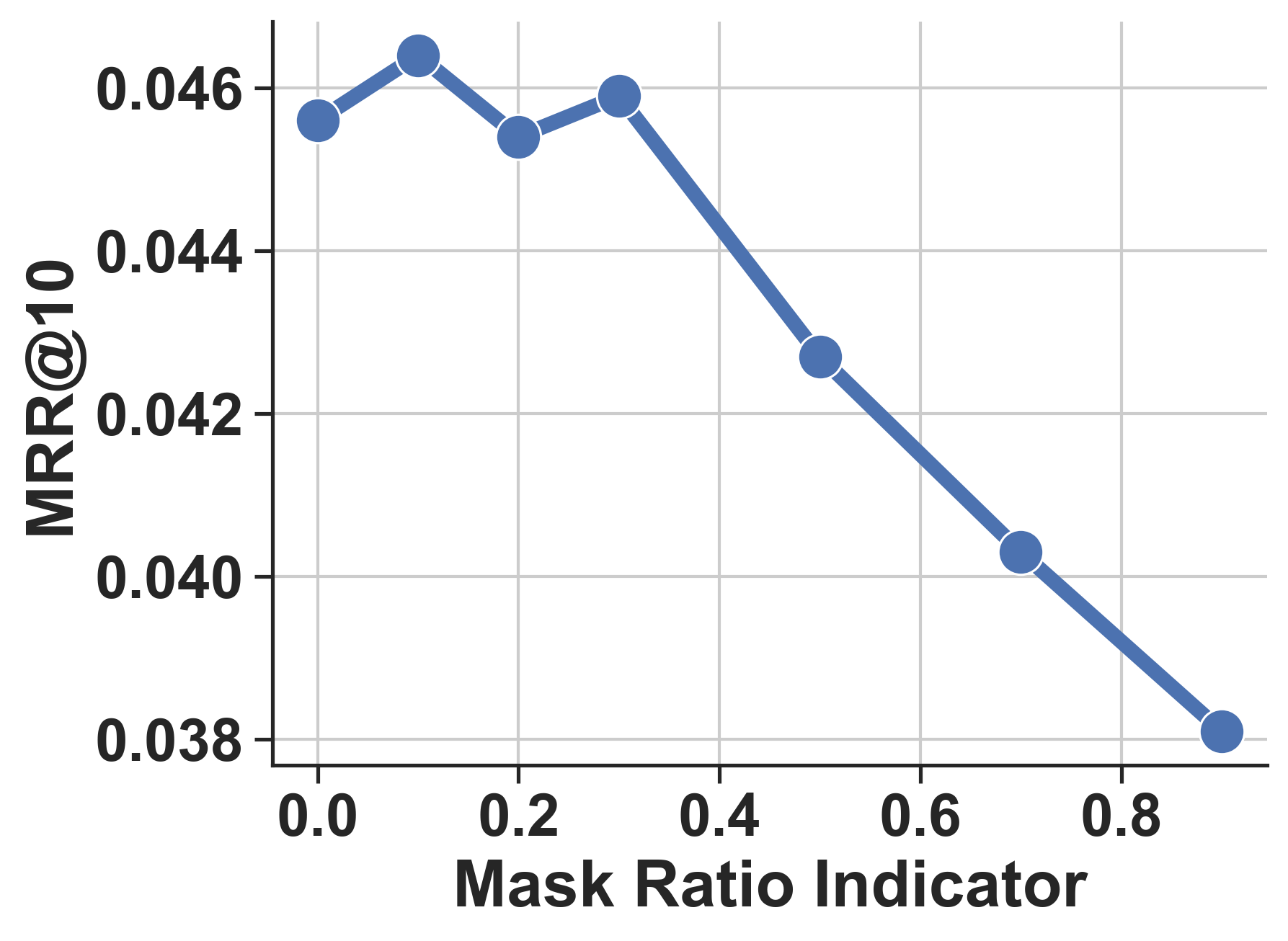}}}
{\subfigure[KuaiRand - MRR@10]
{\includegraphics[width=0.24\linewidth]{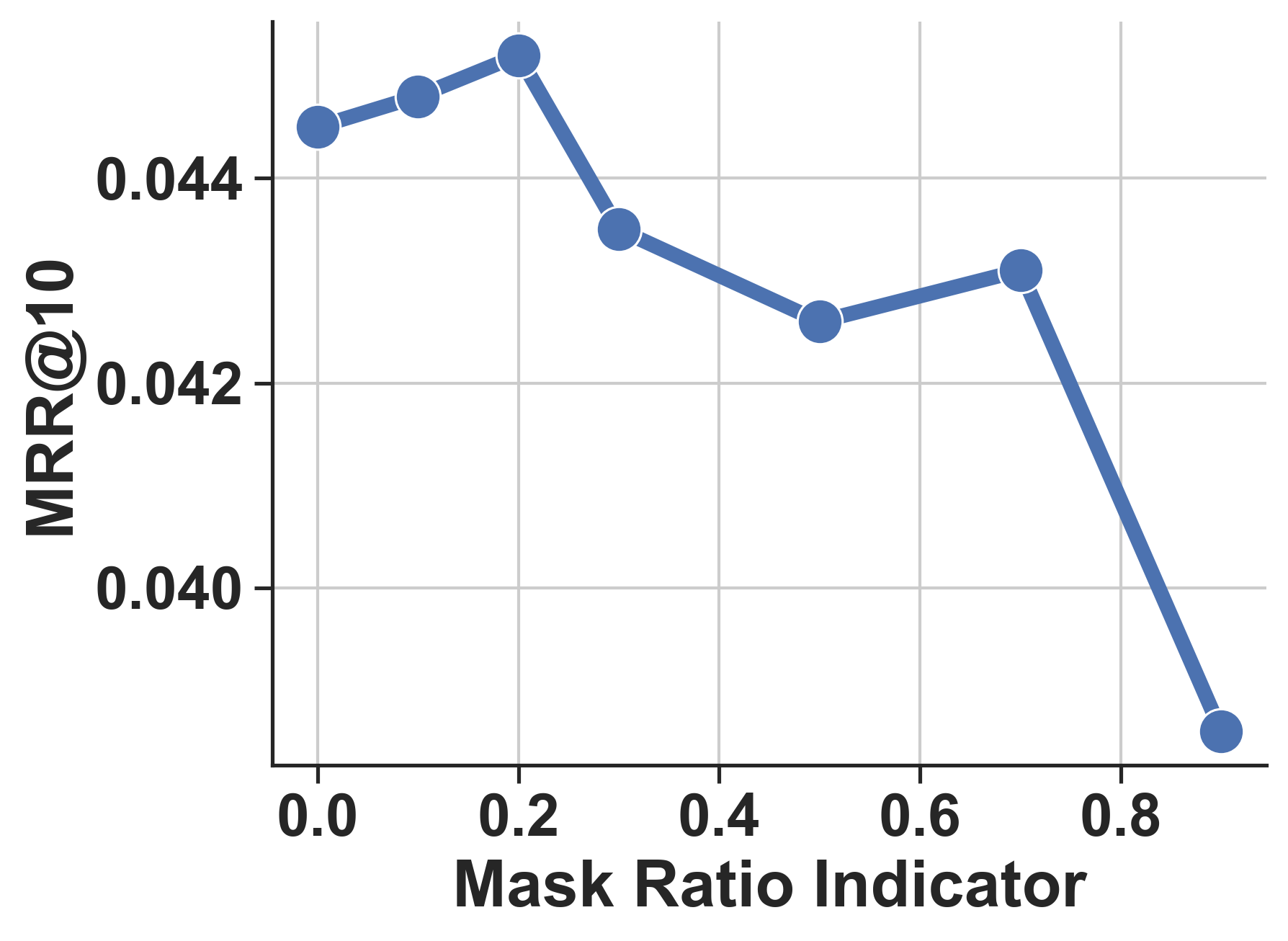}}}

\vspace{1em} 

{\subfigure[ML1M - HR@10]
{\includegraphics[width=0.24\linewidth]{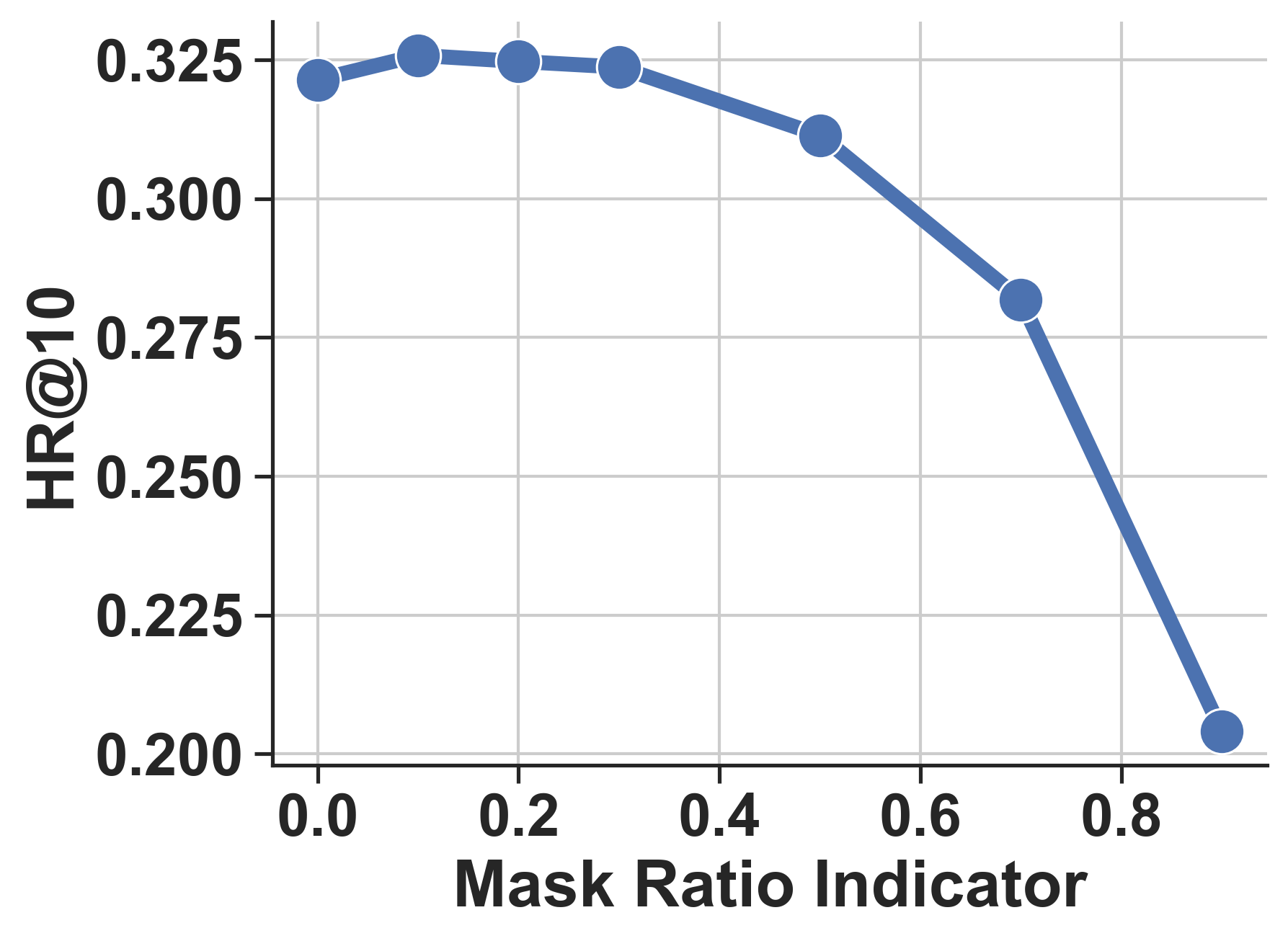}}}
{\subfigure[Beauty - HR@10]
{\includegraphics[width=0.24\linewidth]{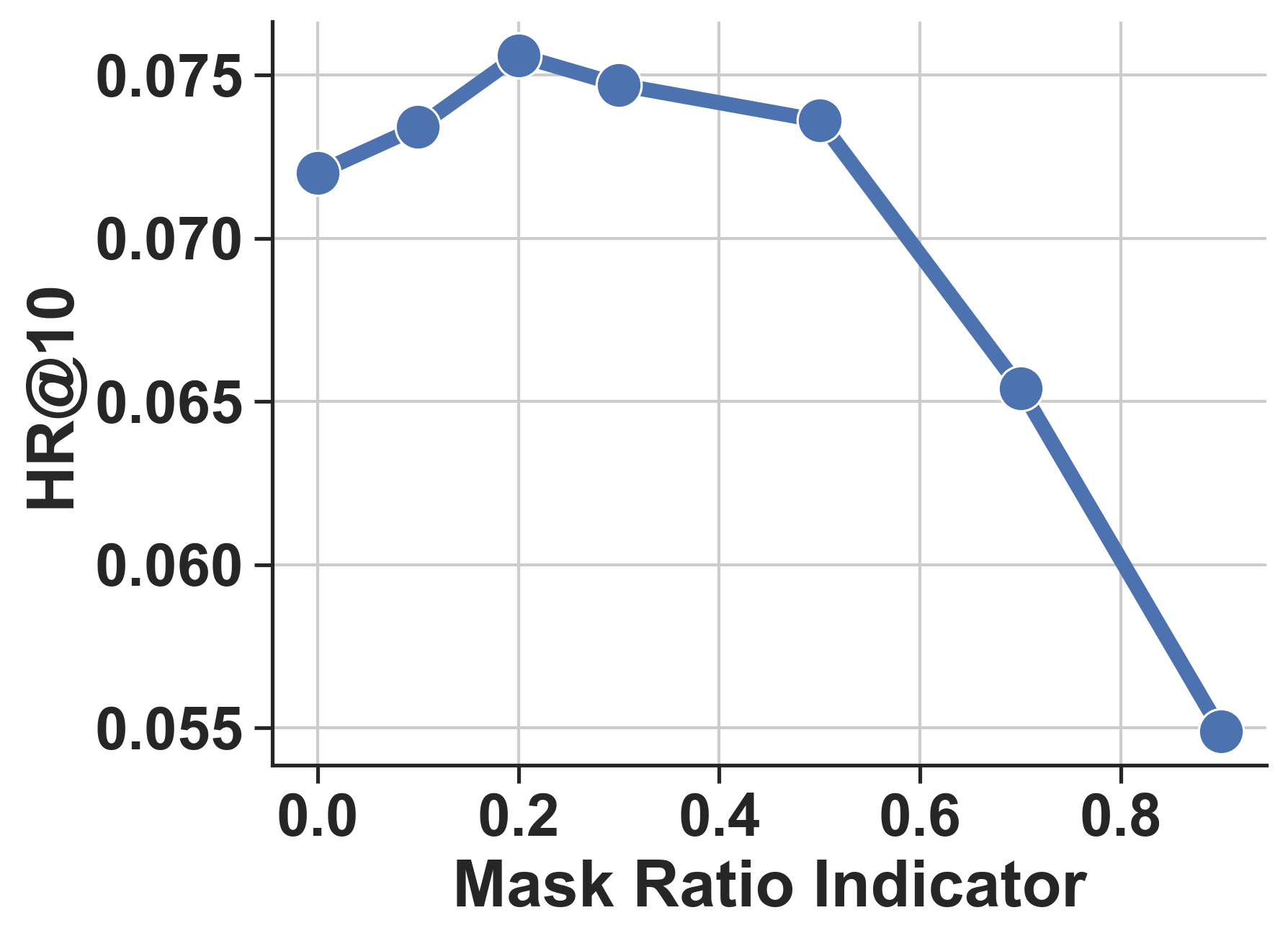}}}
{\subfigure[Games - HR@10]
{\includegraphics[width=0.24\linewidth]{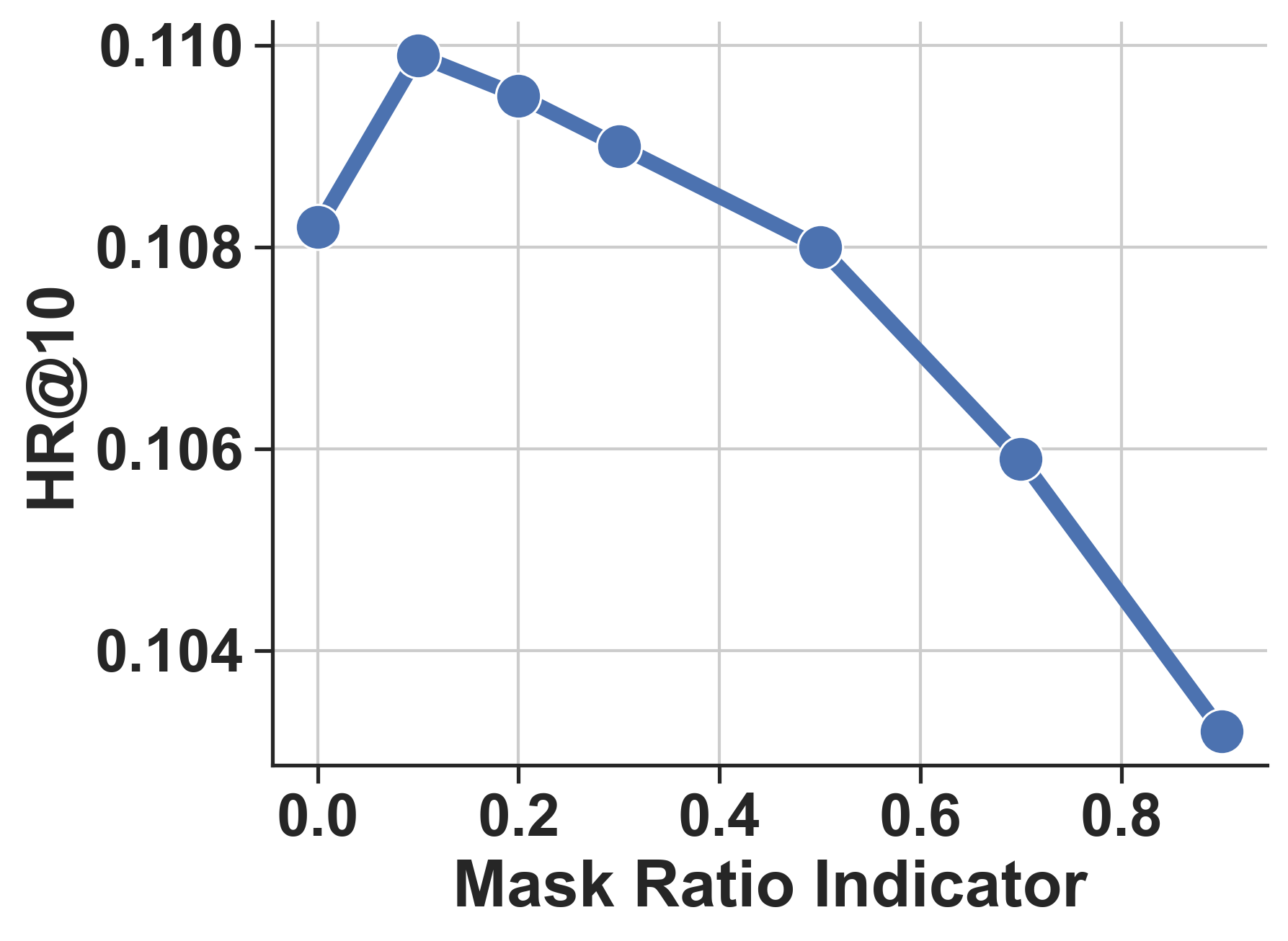}}}
{\subfigure[KuaiRand - HR@10]
{\includegraphics[width=0.24\linewidth]{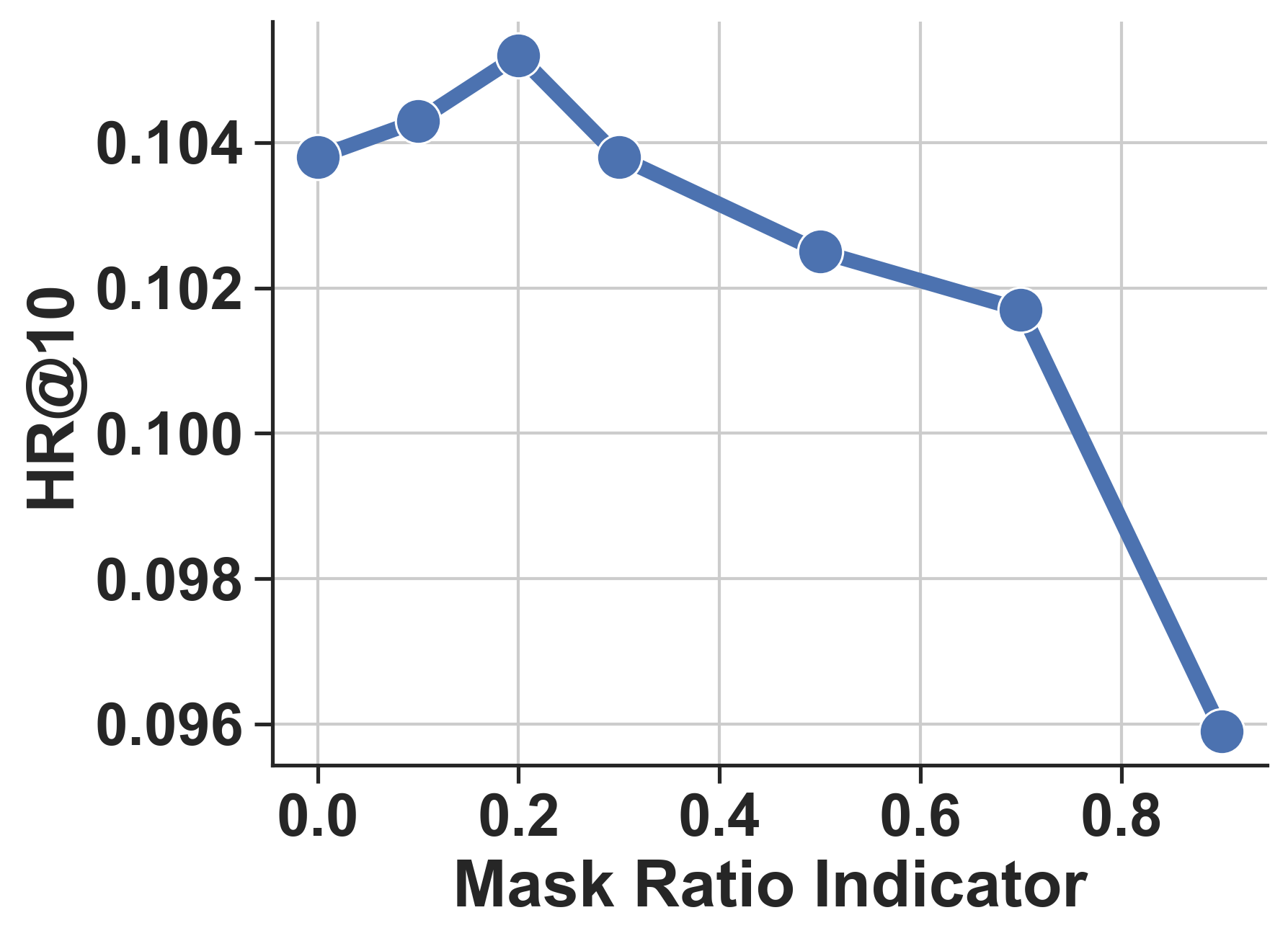}}}
\caption{The effect of mask ratio indicator $\rho$ under NDCG@10, MRR@10, HR@10, separately.}\label{fig:rho_ndcg}
\end{figure*}

This hyperparameter determines the fraction of user-item interaction sequences to be masked.
Figure~\ref{fig:rho_ndcg} show the corresponding performance change of \ourname{} on NDCG@10, MRR@10, HR@10 over $\rho$, separately.
We can find that introducing a small mask ratio brings performance improvements on most metrics.
In most cases, the recommendation performance of our proposed method improves when $\rho\leq 0.2$.
The experimental results also reveal that the recommendation performance degrades as the maski ratio increases, especially when $\rho\geq 0.4$, suggesting excessive mask ratio should be avoided. 
Thus, the ideal mask ratio should take into account the balance between increasing sample diversity to enhance model robustness and preventing excessive information from being masked.
Across the four datasets, namely ML1M, Beauty, Games, and Kuairand, the optimal values $\rho_{best}$ vary: specifically, they are 0.1, 0.2, 0.1, and 0.2, respectively.

\subsubsection{\textbf{Effect of Max Sequence Length $L$}}
\begin{figure*}[htbp]
\centering
{\subfigure[ML1M - NDCG@10]
{\includegraphics[width=0.32\linewidth]{{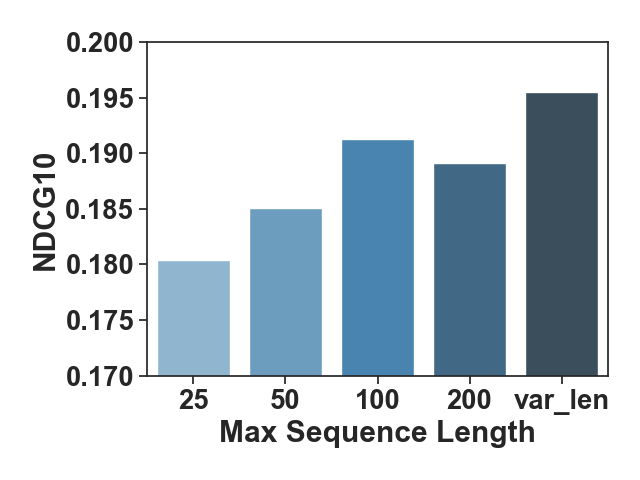}}}}
{\subfigure[ML1M - MRR@10]
{\includegraphics[width=0.32\linewidth]{{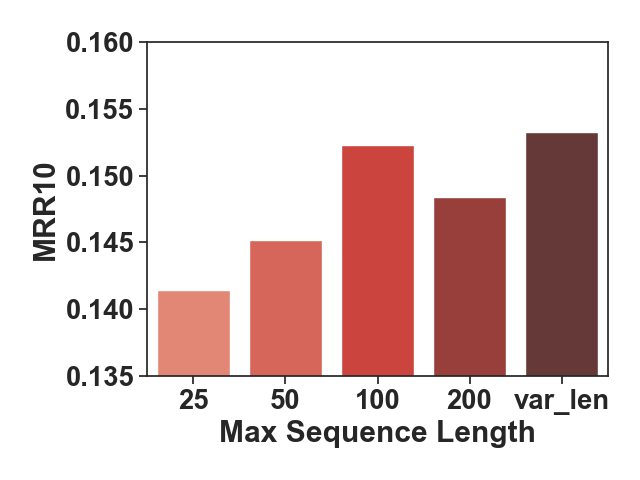}}}}
{\subfigure[ML1M - HR@10]
{\includegraphics[width=0.32\linewidth]{{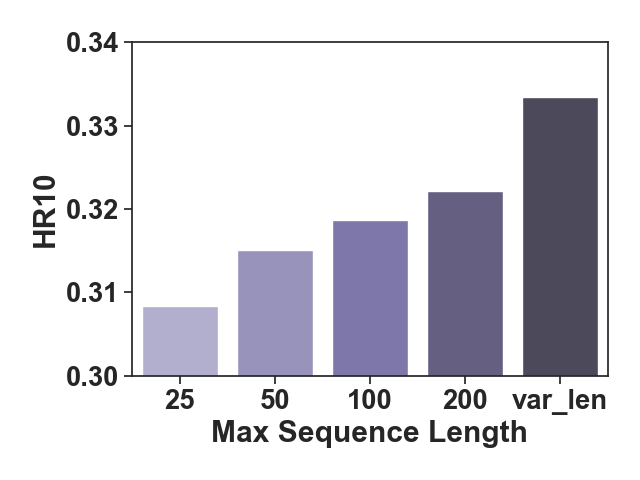}}}}

\vspace{1em} 

{\subfigure[KuaiRand - NDCG@10]
{\includegraphics[width=0.32\linewidth]{{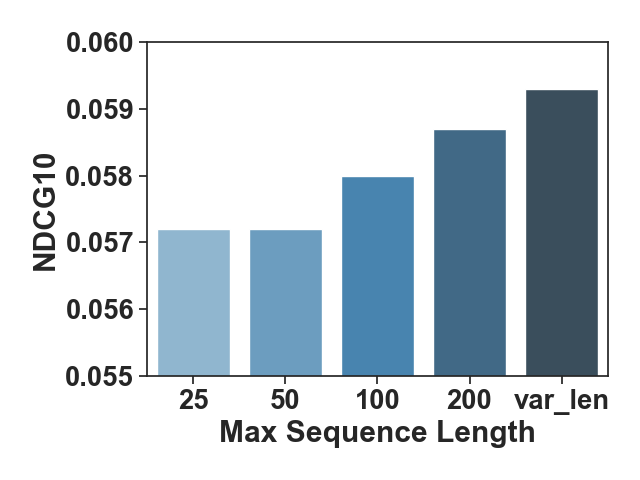}}}}
{\subfigure[KuaiRand - MRR@10]
{\includegraphics[width=0.32\linewidth]{{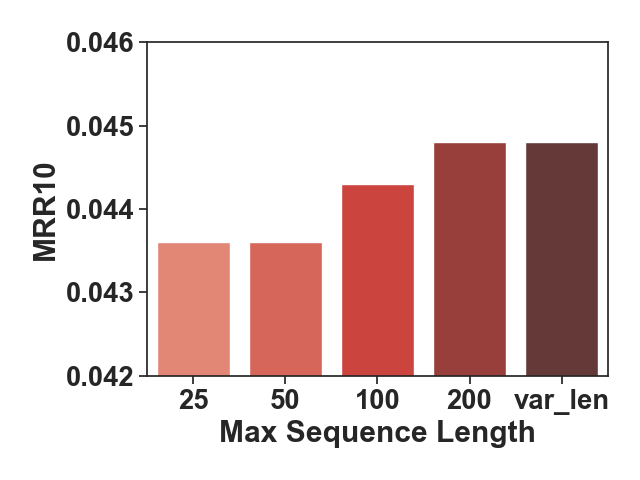}}}}
{\subfigure[KuaiRand - HR@10]
{\includegraphics[width=0.32\linewidth]{{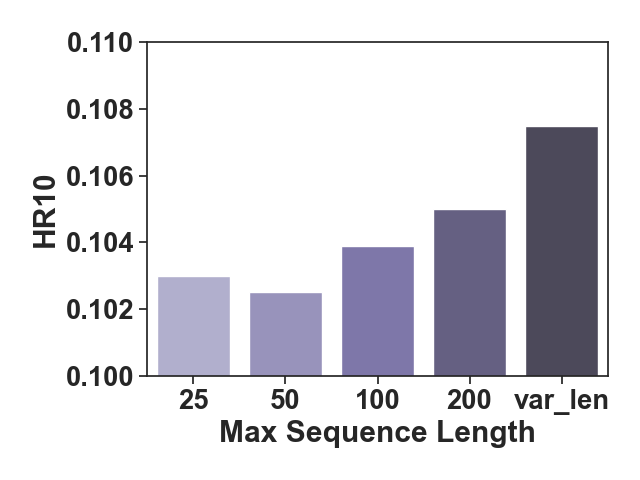}}}}
\caption{The effect of max sequence length $L$ on the ML1M and KuaiRand datasets.}\label{fig:L}
\end{figure*}

To evaluate whether extending sequence lengths benefits our proposed method, we explore hyperparameter values $L$ within the range of {25, 50, 100, 200} for \ourname{} on the ML1M and KuaiRand datasets.
As illustrated in Figure~\ref{fig:L}, our model demonstrates progressively improved performance for sequential recommendations on both datasets as the maximum sequence length increases.
Notably, by leveraging all user interaction histories and extracting valuable information through the sophisticated selection mechanism introduced by Mamba, our approach excels in scenarios with variable-length inputs.
These results underscore the exceptional performance of \ourname{} in long-tail sequential recommendation tasks.

\section{Related Work}
\label{sec:relatedwork}

\subsection{Sequential Recommendation}
In general, existing sequential recommendation studies can be divided into two categories~\cite{fang2020deep}: 

\noindent 1) \textbf{Traditional methods} employ conventional data mining techniques to obtain the users' preferences from their historical interactions~\cite{davidson2010youtube,he2017translation,twardowski2016modelling,liu2023linrec,zhou2022filter}. 
For example, 
as a typical Markov chains-based method, TransRec~\citet{he2017translation} introduces a novel unified method to model third-order interactions in large-scale sequential prediction tasks by embedding items into a transition space and leveraging translation vectors to capture complex user-item and item-item relationships.
Moreover, \citet{he2016fusing} integrates similarity-based approaches with Markov Chains to tackle the sparsity challenges in sequential recommendation tasks.
\citet{twardowski2016modelling} propose a matrix factorization-based sequential RecSys to capture user behavior dependencies without explicit user identification, utilizing user activity within single sessions as sequences of events to provide recommendations.

\noindent 2) \textbf{Deep learning-based methods} leverage the powerful representation capabilities of deep neural networks to model the users' preferences based on the user-item interaction history. 
For example, 
\citet{tang2018personalized} propose a Convolutional Sequence Embedding Recommendation Model for top-N sequential recommendation by embedding users' interaction sequence into an "image" into the time and latent space and using convolutional filters to capture the sequential patterns. 
\citet{zhang2019next} utilize the self-attention mechanisms to estimate the weights for each item within the users' historical behavior to model the users' preferences for a next-item recommendation. 
\citet{chen2018sequential} systematically investigates the impact of temporal information in sequential recommendations by uncovering and incorporating two fundamental temporal patterns of user behaviors, `absolute time patterns' and `relative time patterns', into the memory-augmented neural networks to model user dynamic preferences effectively. 
\citet{tang2019towards} believes that various model architectures excel at capturing different time horizons; some are effective for short-term dependence, while others are better suited for long-term dependence. They propose a hybrid model that incorporates a learned gating mechanism capable of combining different models to effectively manage both short-term and long-term dependencies.
Despite effectiveness, most deep learning-based methods still have an intrinsic limitation on capturing long-range dependencies in user-item interactions:
RNN-, CNN-, and GNN-based methods exhibit inferior performance stemming from memory compression and limited receptive field, while Attention-based methods encounter time-consuming training and inference due to their quadratic complexity with respect to sequence length.

\subsection{Structured State Space Models (Mamba)}
Due to Mamba's powerful sequential modeling capabilities and computational efficiency, numerous studies have applied it across various domains in real-life applications~\cite{qu2024survey}. 
For instance, 
\citet{lieber2024jamba} proposes a cutting-edge large language model called Jamba that combines Transformer and Mamba layers in a hybrid mixture-of-experts architecture, resulting in a high-performance, memory-efficient model capable of handling long-context sequences effectively. 
\citet{ruan2024vm} introduces Vision Mamba UNet (VM-UNet), a novel U-shape architecture model for medical image segmentation that leverages State Space Models (SSMs) to address the limitations of CNNs and Transformers by incorporating a Visual State Space (VSS) block to capture extensive contextual information for medical image segmentation. 
\citet{xing2024segmamba} proposes a pioneering 3D medical image Segmentation Mamba model that leverages the efficiency of State Space Models (SSM) to effectively capture long-range dependencies within whole-volume features at various scales, outperforming Transformer-based methods in whole-volume feature modeling even with high-resolution volume features and achieving superior processing speed and efficacy in medical image segmentation tasks. 
\citet{hwang2024hydra} studies a unifying matrix mixer framework for sequence mixers, which includes a wide range of well-known sequence models, such as Transformers and State Space Models (SSMs). Guided by this framework, they propose Hydra, a natural bidirectional extension of Mamba that corresponds to a quasiseparable matrix mixer. Hydra significantly outperforms transformers in non-causal problems.
Motivated by this success, \cite{liu2024bidirectional} develops a Mamba-based model called SIGMA to tackle the challenges of context modeling and short sequence modeling through the introduction of a Partially Flipped Mamba structure, a Dense Selective Gate, and a Feature Extract GRU. These innovations effectively overcome the limitations of the Mamba model, including its unidirectional structure and instability in state estimation for short sequences.
For more details on existing works, please refer to a comprehensive survey of mamba~\cite{qu2024survey}. 
Different from existing Mamba-based recommendation approaches, our study presents an innovative Masked Bidirectional Structured Space State Model, exploring the potential of a pure-Mamba model in sequential recommendation.
\section{Conclusion}
\label{sec:conclusion}
While existing deep learning-based sequential recommendation methods achieve promising prediction performance, they fail to effectively and efficiently capture user preferences with long-tail interactions.
To tackle these challenges, we propose a novel framework, named \ourname{}, based on  Mamba architecture for sequential recommendation.  
To be specific, the proposed \ourname{} consists of two key modules, i.e., a novel strategy for batch processing of masked variable-length item sequences and a Bi-SSD Layer for efficient user behavior sequence modeling.
Specifically, using the bidirectional structure enables a comprehensive information understanding, and the SSD layer equips the model with attention-like learning capabilities and superior computational efficiency.
Through comprehensive experiments on four distinct datasets, we revealed that our model could achieve state-of-the-art sequential recommendation performance while also exhibiting the capacity to achieve fast training and inference.

\bibliographystyle{ACM-Reference-Format}
\bibliography{references/references}

\end{document}